\documentclass[%
 reprint,
superscriptaddress,
nofootinbib,
 amsmath,amssymb,
 aps,
pra,
]{revtex4-1}

\usepackage{graphicx}
\usepackage{epstopdf}
\usepackage{xcolor}
\usepackage{dcolumn}
\usepackage{bm}
\usepackage{hyperref}
\hypersetup{colorlinks,linkcolor=blue,citecolor=blue}

\usepackage[colorinlistoftodos]{todonotes}

\usepackage{epstopdf}
\usepackage[caption=false]{subfig}
\usepackage{braket}
\usepackage{siunitx}
\usepackage{color}
\usepackage{amsmath}
\usepackage{mathtools}
\usepackage{graphicx}
\usepackage{cancel}
\usepackage{floatrow}
\usepackage{tikz}
\usepackage{todonotes}
\usepackage{hyperref}
\usepackage{physics}

\definecolor{blueE}{HTML}{392BB7}
\definecolor{redE}{HTML}{99004c}
\definecolor{lightblueE}{HTML}{58ACFA}
\definecolor{orangeE}{HTML}{FF4000}
\definecolor{greenE}{HTML}{14754E}
\definecolor{blueC}{rgb}{0,0,200/256}
\definecolor{redC}{rgb}{0.70849190843207621, 0.12647636278656021, 0}

\newcommand{\blueline}{\raisebox{2pt}{\tikz{\draw[-,blueC,solid,line width = 2*1pt](0,0) -- (5mm,0);}}}
\newcommand{\redline}{\raisebox{2pt}{\tikz{\draw[-,redC,solid,line width = 2*1pt](0,0) -- (5mm,0);}}}
\newcommand{\blackline}{\raisebox{2pt}{\tikz{\draw[-,black,solid,line width = 2*1pt](0,0) -- (5mm,0);}}}

\newcommand{\zerozero}{\raisebox{2pt}{\tikz{\draw[-,black,solid,line width = 2*1pt](0,0) -- (5mm,0);}}}
\newcommand{\onezero}{\raisebox{2pt}{\tikz{\draw[-,blueE,solid,line width = 2*1pt](0,0) -- (5mm,0);}}}
\newcommand{\twozero}{\raisebox{2pt}{\tikz{\draw[-,blueE,densely dashed,line width = 2*0.8pt](0,0) -- (5mm,0);}}}
\newcommand{\zeroone}{\raisebox{2pt}{\tikz{\draw[-,redE,solid,line width = 2*0.8pt](0,0) -- (5mm,0);}}}
\newcommand{\oneone}{\raisebox{2pt}{\tikz{\draw[-,greenE,densely dashed,line width = 2*0.8pt](0,0) -- (5mm,0);}}}
\newcommand{\twoone}{\raisebox{2pt}{\tikz{\draw[-,lightblueE,densely dashdotted,line width = 2*0.8pt](0,0) -- (5mm,0);}}}
\newcommand{\zerotwo}{\raisebox{2pt}{\tikz{\draw[-,redE,densely dashed,line width = 2*0.8pt](0,0) -- (5mm,0);}}}
\newcommand{\onetwo}{\raisebox{2pt}{\tikz{\draw[-,orangeE,densely dashdotted,line width = 2*0.8pt](0,0) -- (5mm,0);}}}
\newcommand{\twotwo}{\raisebox{2pt}{\tikz{\draw[-,greenE,densely dotted,line width = 2*0.8pt](0,0) -- (5mm,0);}}}
\newcommand{\nip}{n_{ip}}
\newcommand{\nop}{n_{op}}
\newcommand{\sip}{s_{ip}}
\newcommand{\sop}{s_{op}}

\newcommand{\Mg}{$^{24}$Mg$^+$ }
\newcommand{\Ca}{$^{40}$Ca$^+$ }
\newcommand{\MgH}{$^{24}$MgH$^+$ }
\newcommand{\Mgu}{$^{24}\text{Mg}^+$}
\newcommand{\MgHu}{$^{24}\text{MgH}^+$}

\newcommand{\ir}{i_r}

\newcommand{\et}{e_t}
\newcommand{\er}{e_r}
\newcommand{\gt}{g_t}
\newcommand{\gr}{g_r}
\newcommand{\Rabs}{R_{\text{abs}}}
\newcommand{\Rstim}{R_{\text{stim}}}
\newcommand{\Rspon}{R_{\text{spon}}}

\begin{document}

\title{Unresolved sideband photon recoil spectroscopy of molecular ions}

\author{Emilie H. Clausen}
\affiliation{Department of Physics and Astronomy, Aarhus University, DK-8000 Aarhus C, Denmark}
\author{Vincent Jarlaud}
\altaffiliation[New address: ]{LP2N, IOGS, CNRS and Université de Bordeaux, rue François Mitterrand, 33400 Talence, France}
\affiliation{Department of Physics and Astronomy, Aarhus University, DK-8000 Aarhus C, Denmark}
\author{Karin Fisher}
\affiliation{Department of Physics and Astronomy, Aarhus University, DK-8000 Aarhus C, Denmark}
\author{Steffen Meyer}
\affiliation{Department of Physics and Astronomy, Aarhus University, DK-8000 Aarhus C, Denmark}
\author{Cyrille Solaro}
\altaffiliation[New address: ]{Laboratoire Kastler Brossel (LKB), Sorbonne University, CNRS, ENS-PSL University, Collège de France, Paris, France}
\email{cyrille.solaro@lkb.upmc.fr}
\affiliation{Department of Physics and Astronomy, Aarhus University, DK-8000 Aarhus C, Denmark}
\author{Michael Drewsen}
\email{drewsen@phys.au.dk}
\affiliation{Department of Physics and Astronomy, Aarhus University, DK-8000 Aarhus C, Denmark}
\affiliation{Center for Complex Quantum Systems, Aarhus University, DK-8000 Aarhus C, Denmark}

\date{\today}

\begin{abstract}
We reflect on the prospect of exploiting the recoil associated with absorption and emission of photons to perform spectroscopy of a single molecular ion. For this recoil to be detectable, the molecular ion is sympathetically cooled by a laser-cooled atomic ion to near their common quantum mechanical ground state within a trapping potential. More specifically, we present a general framework for simulating the expected photon recoil spectra in regimes where either the natural transition linewidth $\Gamma_t$ of the molecular ion or the spectral width $\Gamma_L$ of the exciting light source exceeds the motional frequencies of the two-ion system. To exemplify the framework, we present two complementary cases: spectroscopy of the broad 3s $^2$S$_{1/2}$ - 3p $^2$P$_{3/2}$ electronic transition ($\Gamma_t/2\pi = \SI{41.8}{\mega\hertz}$) of a single \Mg ion at $\lambda=279.6$ nm by a narrow laser source ($\Gamma_L/2\pi \lesssim 1$ MHz) and mid-infrared vibrational spectroscopy of the very narrow $\ket{v=0,J=1}$ - $\ket{v'=1,J'=0}$ transition ($\Gamma_t/2\pi = 2.50 $ Hz) at $\lambda=\SI{6.17}{\micro\meter}$ in the $^1\Sigma^+$ electronic ground state of \MgH by a broadband laser source ($\Gamma_L/2\pi \gtrsim$ 50 MHz). The atomic ion \Mg has been picked to introduce a simple system to make comparisons with experimental results while still capturing most of the physics involved in electronic excitations of molecular ions.
\end{abstract}


\maketitle

\section{Introduction}

In the past two decades, it has become possible to trap and sympathetically cool ensembles of molecular ions in the gas phase to the millikelvin range, where they become part of so-called Coulomb crystals through interactions with simultaneously trapped and laser-cooled atomic ions \cite{Molhave2000}. More recently, single molecular ions have even been sympathetically cooled to microkelvin temperatures by single atomic ions, where the common modes of the strongly coupled two-ion system are close to their quantum mechanical ground states \cite{Gregers,Wan2015,Rugango2015}. The latter scenario constitutes a novel setting for conducting molecular spectroscopy with potentially very high resolution for fundamental structure studies of molecular ions, tests of fundamental physics theories, and quantum technology oriented applications \cite{Wolf2016,Chou2017,Chou2019,Sinhal2020,Najafian2020,Lin2020,Wolf2020}. However, since it is practically impossible to count a single absorbed photon from a light beam, or to detect with high probability a single photon emitted by a single molecular ion, standard absorption and emission spectroscopy cannot be applied efficiently. 
The solution to this problem is photon recoil spectroscopy (PRS) \cite{Schmidt2005,Wan2014,Gebert2015,Schulte2018}, where it is the momentum recoil associated with absorption and emission of individual photons by a single target ion that signals a spectroscopic event through the excitation of the common motion of the two-ion system. Since PRS has been already applied to ultra-precise spectroscopy of atomic ions \cite{Rosenband2008} and should be largely applicable to molecular ions as well, it holds great promises for ultra-high resolution spectroscopy of molecular ions in the near future \cite{Chou2019}. 
PRS is also interesting for a range of other investigations of molecules in the gas phase. 
Specific implementations of this technique could include internal state preparation of molecules \cite{Schmidt2006,Vogelius2006,Wolf2016,Chou2017,Sinhal2020}, e.g. for state to state reaction experiments, or single photon absorption studies of single, complex, molecular ions under well-controlled conditions.
In the latter example, the technique can even be applied in situations where the absorption leads to complete internal energy conversion.

In the following section (Sec. \ref{model}), we will discuss in detail the basics of PRS and present a mathematical framework which can be used to simulate photon recoil spectra in the special cases where the spectra are unresolved with respect to the motional sidebands, either due to naturally occurring broad spectroscopic transitions of linewidth $\Gamma_t$ or due to the linewidth $\Gamma_L$ of the applied light sources. This generic section will serve as the basis for the next section (Sec. \ref{Simu}), where we present simulated results for the spectroscopy of the broad 3s $^2$S$_{1/2}$ - 3p $^2$P$_{3/2}$ electronic transition ($\Gamma_t/2\pi = \SI{41.8}{\mega\hertz}$ \cite{Herrmann2009,Mg_spec_aug}) of a single \Mg ion at $\lambda=279.6$ nm by a narrow laser source ($\Gamma_L/2\pi \lesssim 1$ MHz), and mid-infrared vibrational spectroscopy by a broadband laser source ($\Gamma_L/2\pi \gtrsim$ 50 MHz) of the very narrow $\ket{v=0,J=1}$ - $\ket{v'=1,J'=0}$ transition ($\Gamma_t/2\pi =2.50$ Hz \cite{Hansen,Frank})  at $\lambda=\SI{6.17}{\micro\meter}$ in the $^1\Sigma^+$ electronic ground state of \MgHu. 
The prospects and limitations of unresolved PRS are discussed in Sec. \ref{Discussion} before the conclusion in Sec. \ref{Conclusion}.

\section{Model for unresolved sideband PRS}\label{model}

\begin{figure}          
\includegraphics[width = 1\textwidth]{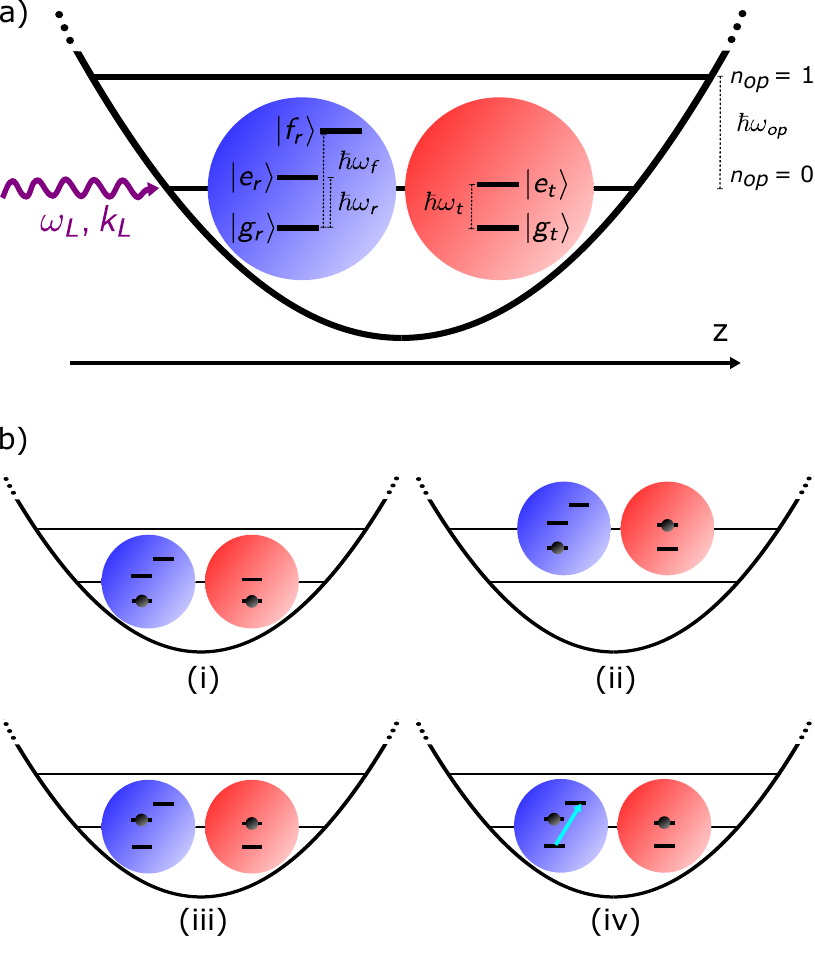}
\caption{a) A single spectroscopic target ion (red) is trapped together with one readout atomic ion (blue) in a linear Paul trap. Doppler cooling followed by sideband cooling on the readout ion ensures that the two-ion Coulomb crystal is in the motional ground state along $z$. The simplified internal structure of each ion is also shown: the target ion has a simple two-level structure and the readout ion has a three-level structure with two excited states, one long-lived ($\ket{e_r}$) and one short-lived ($\ket{f_r}$) used for fluorescence detection. A light field of frequency $\omega_L$ propagating along $z$ interacts with one of the two ions only.
b) Schematic of PRS. (i) The two-ion system is initialized in its motional ground state along $z$ with respect to e.g. the out-of-phase mode of motion. The readout ion is prepared in its electronic ground state and the molecular ion in its internal target state. (ii) After a resonant BSB spectroscopy pulse on the target ion, the two-ion system is transferred to the first excited state $n_{op} = 1$. (iii) After a resonant RSB pulse on the readout ion, the two-ion system is transferred back to the motional ground state, and the readout ion is in the $|e_r\rangle$ state. (iv) When exposed to light resonant with the $|g_r\rangle - |f_r\rangle$ transition, the readout ion does not fluoresce, reflecting a successful excitation of the target ion by the BSB pulse before step (ii).}
\label{expQLS}
\end{figure}

The idea of exploiting the photon recoil associated with absorption and emission in connection with spectroscopy was first devised by the group of Nobel laureate Prof. Wineland. 
It was devised with the prospect of developing optical atomic clocks based on single atomic ions with suitably narrow optical transitions, but lacking transitions for direct laser cooling \cite{WinelandProc,Schmidt2005}. 
In this case, spectroscopy is carried out by trapping a single spectroscopic target ion together with a single atomic ion that can be sideband cooled. Via the Coulomb interaction between the ions, the two-ion system can be brought to the quantum mechanical ground state with respect to one or more motional modes. 
In the original paper, the authors consider a so-called resolved sideband scenario where the two motional mode angular frequencies corresponding to the in-phase mode ($\omega_{ip}$) and out-of-phase mode ($\omega_{op}$) are significantly larger than both the transition linewidth and the spectral width of the spectroscopy laser. In this situation, it is possible to selectively address both the spectroscopy ion (target ion) and the sideband-cooling ion (readout ion) with lasers tuned resonantly to either carrier or specific motional sideband transitions. 
Fig. \ref{expQLS} illustrates one of the simplest specific implementations of PRS capturing the elements important to the present paper. This procedure starts by (i) initializing the two-ion system in the quantum mechanical ground state of at least one of the two motional modes along the axis defined by the two ions (e.g. the out-of-phase mode), the readout atomic ion in its electronic ground state, and the molecular ion in its internal target state for the spectroscopy. Next, the target molecular ion is exposed to a light pulse expected to be resonant with the blue sideband (BSB) of the spectroscopy transition (i.e. $\omega_L=\omega_t+\omega_{op}$, where $\omega_L$, $\omega_t$ are the angular frequencies of the light and of the target transition respectively). 
If the correct interaction time $\tau_{\text{spec}}$ is chosen, the resonant BSB pulse leads to a full excitation of the target ion to $\ket{e_t}$ (see the more complete theoretical description below), and the two-ion system is transferred to the state depicted in (ii), where the out-of-phase mode is now in its first excited state ($n_{op}=1$).
This motional excitation can be monitored by addressing the readout atomic ion with a light pulse resonant with the red sideband (RSB) of the narrow sideband cooling transition ($\omega_L=\omega_r-\omega_{op}$, where $\omega_r$ is the transition frequency) for a time $\tau_r$ corresponding to a full excitation to the $\ket{e_r}$ state at the expense of the motional excitation (iii). In the final step (iv), the readout ion is exposed to light resonant with a closed fast fluorescing transition $\ket{g_r} - \ket{f_r}$. This leads to the emission of several photons at the frequency $\omega_f$ if the readout ion was in the $\ket{g_r}$ state, versus no photon emission if in the $\ket{e_r}$ state. Since we assumed in step (iii) to have brought the readout ion to the $\ket{e_r}$ state, we expect no fluorescence in the last step (iv) if the pulse applied to the target molecular ion was indeed excited by the first BSB pulse.
Conversely, if the target ion was not excited on the BSB, the readout ion would stay in the $\ket{g_r}$ state after step (iii), and fluorescence light would be emitted during the final step (iv). Hence, through repetition of the PRS sequence (i)-(iv) for different values of $\omega_L$ when addressing the target ion, the total fluorescence signal from the readout atomic ion will reflect the excitation probability of the target molecular ion and thus produce a spectroscopy signal.

In the next sub-section \ref{ResolvedPRS} we will present a mathematical framework which can be applied to simulate the expected spectroscopy signal for the resolved sideband scenario presented above. On the basis of this we will in sub-section \ref{UnresolvedPRS} establish a general framework to also describe the expected signal in unresolved sideband regimes.

\subsection{Hamiltonian dynamics relevant to PRS in the resolved sideband regime}\label{ResolvedPRS}

In the following we consider the situation presented in Fig. \ref{expQLS} with the two ions, the target and the readout ion, both confined in a trap along the $z$-axis (e.g. the rf-free axis of a linear rf trap) thanks to suitably high trapping frequencies in the perpendicular plane. We assume that the internal states of the two ions are the same as the ones defined in Fig. \ref{expQLS}. The target ion has a simple two-level structure $\lbrace \ket{g_t}, \ket{e_t}\rbrace$ which leads to the internal state Hamiltonian
\begin{align}
\hat{H}_{t}=\frac{\hbar\omega_{t}}{2}\left(\ket{e_t}\bra{e_t}-\ket{g_t}\bra{g_t}\right)
\end{align}
with $\hbar\omega_t=E_{e,t}-E_{g,t}$, where $E_{g,t}$ and $E_{e,t}$ are the energies of the two states. The readout ion has the three-level structure $\lbrace \ket{g_r}, \ket{e_r}, \ket{f_r}\rbrace$. It reduces to the two levels $\ket{g_r}$ and $\ket{e_r}$ when considering only the Hamiltonian evolution of the PRS sequence before readout (steps (i) to (iii), Fig. \ref{expQLS}). The internal state Hamiltonian is thus
\begin{align}
\hat{H}_{r}=\frac{\hbar\omega_{r}}{2}\left(\ket{e_r}\bra{e_r}-\ket{g_r}\bra{g_r}\right)
\end{align}
with $\hbar\omega_r=E_{e,r}-E_{g,r}$, where $E_{g,r}$ and $E_{e,r}$ are the energies of the two states. In addition to the two ions' internal states, the external motional states of the two-ion system along the $z$-axis can be described by the following Hamiltonian
\begin{align}
\hat{H}_z=\hbar\omega_{ip}\bigg(\hat{a}_{ip}^{\dagger}\hat{a}_{ip}+\frac{1}{2}\bigg)+\hbar\omega_{op}\bigg(\hat{a}_{op}^{\dagger}\hat{a}_{op}+\frac{1}{2}\bigg)
\end{align}
where we have introduced the standard harmonic oscillator ladder operators $\hat{a}_{ip}^{\dagger}$, $\hat{a}_{ip}$ and $\hat{a}_{op}^{\dagger}$, $\hat{a}_{op}$ for the in-phase (IP) and out-of-phase (OP) modes respectively. The associated mode angular frequencies, for singly charged ions, are given by \cite{Morigi2001,Drewsen2004}
\begin{align}
\begin{split}
\omega_{ip}=\omega_z\bigg(1+\frac{1}{\mu}-\sqrt{1-\frac{1}{\mu}+\frac{1}{\mu^2}}\bigg)^{1/2}\\
\omega_{op}=\omega_z\bigg(1+\frac{1}{\mu}+\sqrt{1-\frac{1}{\mu}+\frac{1}{\mu^2}}\bigg)^{1/2}
\end{split}
\label{modesFreqs}
\end{align}
with $\mu=m_t/m_r$ being the ion mass ratio and $\omega_z$ the motional angular frequency of the readout ion if it were alone in the trap, all trapping parameters being equal. Essential for PRS is the Hamiltonian describing the light induced interaction between the internal and external degrees of freedom of the ions. We assume here that the light field can be approximated by a monochromatic plane wave with angular frequency $\omega_L$ travelling along the $z$-axis. Furthermore, we assume that the difference between the two transition angular frequencies $\omega_t$ and $\omega_r$ is large enough that a light field close to resonance with one of the ions will not perturb the other. In this case, one can write the Hamiltonian for the light interaction as:
\begin{align}
\hat{H}_{int,j}=\hbar\Omega_{0,j}\cos(k_L\hat{z}_j-\omega_L t)(\ket{e_j}\bra{g_j}+\ket{g_j}\bra{e_j})
\end{align}
where $j\in\lbrace r,t\rbrace$ and $\Omega_{0,j}$ is the vacuum Rabi angular frequency associated with the particular light field and oscillator strength and $\vec{k}_L$ is the wave vector of the laser field.
The coupling of the ions' internal and external degrees of freedom appears through the position operator $\hat{z}_j$ of the interaction Hamiltonian. This operator can be written for both ions in terms of the ladder operators as \cite{wineland1998,Leibfried2003,Wubbena2012}:
\begin{align}
\hat{z}_j=\vert b_{ip,j}\vert&\sqrt{\frac{\hbar}{2m_j\omega_{ip}}}(\hat{a}_{ip}^{\dagger}+\hat{a}_{ip}) \nonumber \\
&+\vert b_{op,j}\vert\sqrt{\frac{\hbar}{2m_j\omega_{op}}}(\hat{a}_{op}^{\dagger}+\hat{a}_{op})
\end{align}
where $b_{ip,j}$ and $b_{op,j}$ are the components of the eigenvectors (in a mass-weighted space) for the IP and OP modes respectively. They can be written for the readout and target ions as:
\begin{align}
\begin{split}
b_{ip/op,r} = \frac{\beta_{ip/op}}{\sqrt{1 + \beta_{ip/op}^2}}\\
b_{ip/op,t} = \frac{1}{\sqrt{1 + \beta_{ip/op}^2}}
\end{split}
\end{align}
with
\begin{align}
\begin{split}
\beta_{ip} &= \frac{-\mu + 1 + \sqrt{\mu^2 - \mu + 1}}{\sqrt{\mu}} \\
\beta_{op} &= \frac{-\mu + 1 - \sqrt{\mu^2 - \mu + 1}}{\sqrt{\mu}}.
\end{split}
\end{align}
Introducing the Lamb-Dicke parameters (LDPs) \cite{Morigi2001}
\begin{align}
\begin{split}
&\eta_{ip,j}=\vec{k}_L\cdot \vec{e}_z\vert b_{ip,j}\vert\sqrt{\frac{\hbar}{2m_j\omega_{ip}}} \quad \\
&\eta_{op,j}=\vec{k}_L\cdot \vec{e}_z\vert b_{op,j}\vert\sqrt{\frac{\hbar}{2m_j\omega_{op}}},
\label{Lamb}
\end{split}
\end{align}
where $\vec{e}_z$ is a unit vector along $z$, we can rewrite the interaction Hamiltonian as:
\begin{widetext}
\begin{align}
\hat{H}_{int,j}=\hbar\Omega_{0,j}\cos\big(\eta_{ip,j}(\hat{a}_{ip}^{\dagger}+\hat{a}_{ip})+\eta_{op,j}(\hat{a}_{op}^{\dagger}+\hat{a}_{op})-\omega_L t\big)\big(\ket{e_j}\bra{g_j}+\ket{g_j}\ket{e_j}\big).
\end{align}
The full Hamiltonian is given by
\begin{align}
\hat{H}_{tot}=\hat{H}_r+\hat{H}_t+\hat{H}_z+\hat{H}_{int,j}\equiv\hat{H}_0+\hat{H}_{int,j}.
\end{align}
In order to investigate the dynamics of the interaction between the internal and external degrees of freedom, it is convenient to work in the interaction picture through the transformation
\begin{align}
\hat{H}_I=e^{i\hat{H}_0t/\hbar}\hat{H}_{int,j}e^{-i\hat{H}_0t/\hbar}
\end{align}
which, due to the commutativity between the internal and external state operators, leads to
\begin{align}
\begin{split}
\hat{H}_I=\hbar\Omega_{0,j}\Big(e^{i\omega_jt}\ket{e_j}\bra{g_j}+e^{-i\omega_jt}\ket{g_j}\bra{e_j}\Big) e^{i\hat{H}_zt/\hbar}\cos\big(\eta_{ip,j}(\hat{a}_{ip}^{\dagger}+\hat{a}_{ip})+\eta_{op,j}(\hat{a}_{op}^{\dagger}+\hat{a}_{op})-\omega_L t\big)e^{-i\hat{H}_zt/\hbar}.
\end{split}
\end{align}
Performing the rotating wave approximation, i.e. keeping time dependent terms containing $\omega_j-\omega_L$ and not the ones containing $\omega_j + \omega_L$, one arrives at
\begin{align}
\hat{H}_I=\frac{\hbar}{2}\Omega_{0,j}\ket{e_j}\bra{g_j}e^{i\hat{H}_zt/\hbar}e^{i(\eta_{ip,j}(\hat{a}_{ip}+\hat{a}_{ip}^{\dagger})+\eta_{op,j}(\hat{a}_{op}+\hat{a}_{op}^{\dagger})-\delta_j t)}e^{-i\hat{H}_zt/\hbar}+h.c.,
\end{align}
where we have introduced the laser detuning $\delta_j = \omega_L-\omega_j$. By introducing the time scaled ladder operators
\begin{align}
\begin{split}
\tilde{\hat{a}}_{ip}=\hat{a}_{ip}e^{-i\omega_{ip}t}\quad &,\quad \tilde{\hat{a}}_{ip}^{\dagger}=\hat{a}_{ip}^{\dagger}e^{i\omega_{ip}t}\\
\tilde{\hat{a}}_{op}=\hat{a}_{op}e^{-i\omega_{op}t}\quad &,\quad \tilde{\hat{a}}_{op}^{\dagger}=\hat{a}_{op}^{\dagger}e^{i\omega_{op}t}
\end{split}
\end{align}
and using the Baker-Campbell-Haussdorff formula for decomposing the exponential term \cite{Magnus1954}, one can re-express the interaction Hamiltonian as:
\begin{align}
\hat{H}_I=\frac{\hbar\Omega_{0_j}}{2}\ket{e_j}\bra{g_j}e^{-i\delta_j t}e^{-\eta_{ip,j}^2/2}e^{-\eta_{op,j}^2/2}e^{i\eta_{ip,j}\tilde{\hat{a}}_{ip}^{\dagger}}e^{i\eta_{ip,j}\tilde{\hat{a}}_{ip}}e^{i\eta_{op,j}\tilde{\hat{a}}_{op}^{\dagger}}e^{i\eta_{op,j}\tilde{\hat{a}}_{op}}+h.c.
\end{align}
\end{widetext}
The specific quantum dynamics depends on which ion we address with the light field. Next, one has to solve the Schr{\"o}dinger equation of motion
\begin{align}
i\hbar\frac{\partial}{\partial t}\ket{\psi(t)}=\hat{H}\ket{\psi(t)}
\label{schrodinger}
\end{align}
in the state-basis of $\ket{i_r,i_t,n_{ip},n_{op}}$, where $\ket{i_{j}}$ indicates the internal state $i\in\lbrace g,e\rbrace$ of the ion $j\in\lbrace r,t\rbrace$. In the case where the readout ion is addressed by the light field, we obtain the following interaction matrix elements \cite{Wineland1979}
\begin{align}
\langle e_r, i_t, &~n_{ip}',n_{op}'|\hat{H}_I|g_r,i_t,n_{ip},n_{op}\rangle= \nonumber \\
& \frac{\hbar\Omega_{0,r}}{2}e^{i(s_{ip}\omega_{ip}+s_{op}\omega_{op}-\delta_r)t}
e^{-\eta_{ip,r}^2/2}e^{-\eta_{op,r}^2/2} \nonumber \\
&\times\sum\limits_{m_{ip}=0}^{n_{ip}^<}
 \frac{(i\eta_{ip,r})^{2m_{ip}+\vert s_{ip}\vert}\sqrt{n_{ip}^<!n_{ip}^>!}}{m_{ip}!(m_{ip}+\vert s_{ip}\vert)!(n_{ip}^<-m_{ip})!} \nonumber \\
& \times \sum\limits_{m_{op}=0}^{n_{op}^<} \frac{(i\eta_{op,r})^{2m_{op}+\vert s_{op}\vert}\sqrt{n_{op}^<!n_{op}^>!}}{m_{op}!(m_{op}+\vert s_{op}\vert)!(n_{op}^<-m_{op})!}
\label{matrixElement}
\end{align}
where $s_{ip/op}\equiv n'_{ip/op}-n_{ip/op}$ is the sideband order, $n_{ip/op}^<\equiv\min\lbrace n_{ip/op},n_{ip/op}'\rbrace$ and $n_{ip/op}^>\equiv\max\lbrace n_{ip/op},n_{ip/op}'\rbrace$. In a slightly more compact form, we can rewrite this expression as
\begin{align}
\bra{e_r,i_t,n_{ip}',n_{op}'}&\hat{H}_I\ket{g_r,i_t,n_{ip},n_{op}}= \nonumber \\
 \frac{\hbar\Omega_{0,r}}{2} & e^{i(s_{ip}\omega_{ip}+s_{op}\omega_{op}-\delta_r)t} \nonumber \\
&~\times \xi(\eta_{ip,r},\eta_{op,r},n_{ip},n_{op},s_{ip},s_{op})
\label{compactMatrixElement}
\end{align}
with
\begin{align}
\xi(\eta_{ip,r},& \eta_{op,r},n_{ip},n_{op},s_{ip},s_{op})= \nonumber \\
&\left(e^{-\eta_{ip,r}^2/2} (i\eta_{ip,r})^{\vert\sip\vert}\sqrt{\frac{n_{ip}^<!}{n_{ip}^>!}}L_{n_{ip}^<}^{\vert\sip\vert}(\eta_{ip,r}^2)\right) \nonumber  \\
\times &\left(e^{-\eta_{op,r}^2/2} (i\eta_{op,r})^{\vert\sop\vert}\sqrt{\frac{n_{op}^<!}{n_{op}^>!}}L_{n_{op}^<}^{\vert\sop\vert}(\eta_{op,r}^2)\right)
\label{scalingFactor}
\end{align}
where $L_{n_{ip}^<}^{\vert\sip\vert}(\eta_{ip,r}^2)$ and $L_{n_{op}^<}^{\vert\sop\vert}(\eta_{op,r}^2)$ are the generalized Laguerre polynomials. 
One can see from Eq. (\ref{compactMatrixElement}) that, for the monochromatic situation considered now, dynamics involving transitions with specific changes in motional quantum numbers can be efficient by tuning the laser field to be resonant with a specific sideband, i.e.,
\begin{align}
s_{ip}\omega_{ip}+s_{op}\omega_{op}-\delta_r=0.
\end{align}
In this case the explicit time dependence in the coupling matrix elements vanishes for the particular type of transitions, while the off-resonant coupling terms to other transitions will generally average to zero. The Rabi angular frequency for a specific sideband and motional state is thus given by \cite{wineland1998}
\begin{align}
\Omega_{n_{ip},n_{op},s_{ip},s_{op}} = \Omega_{0,r}\xi(\eta_{ip,r},\eta_{op,r},n_{ip},n_{op},s_{ip},s_{op}). 
\label{rabi}
\end{align}
The situation becomes particularly simple in the so-called Lamb-Dicke regime where $\eta_{ip/op,r}\sqrt{2\langle n_{ip/op}\rangle+1}\ll 1$. In this case, we can simplify Eq. (\ref{scalingFactor}) to \cite{wineland1998}
\begin{align}
\xi(\eta_{ip,r},&\eta_{op,r},n_{ip},n_{op},s_{ip},s_{op})= \nonumber \\ 
&\Bigg(\frac{\eta_{ip,r}^{\vert\sip\vert}}{\vert \sip\vert!}\sqrt{\frac{\nip^>!}{\nip^<!}}\Bigg)\Bigg(\frac{\eta_{op,r}^{\vert\sop\vert}}{\vert \sop\vert!}\sqrt{\frac{\nop^>!}{\nop^<!}}\Bigg)
\label{scalingLD}
\end{align}
for $s_{ip/op}=0,\pm 1$, and zero otherwise. For resolved sideband PRS, the typical starting point is to have both motional modes cooled to the quantum mechanical ground state. Eq. (\ref{scalingLD}) is then typically a good approximation for simulating the internal and external quantum dynamics before the final unresolved sideband detection addressing the $\ket{g_r} - \ket{f_r}$ transition of the readout ion (step (iv), Fig. \ref{expQLS}). This readout signal corresponds to a projection measurement of the readout ion to its ground state $\ket{g_r,i_t,n_{ip},n_{op}}$. Formally, it is proportional to 
\begin{align}
P_{\ket{g_r}} = \sum_{n_{ip},n_{op},i_t} \vert \braket{\psi_{\text{Ham}}}{g_r,i_t,n_{ip},n_{op}}\vert^2,
\label{proj}
\end{align} 
with $\psi_{\text{Ham}}$ being the wavefunction after the Hamiltonian evolution of steps i) to iii) of the PRS sequence.

\subsection{PRS in the unresolved sideband regime} \label{UnresolvedPRS}

Although PRS was originally developed for ultra-precise spectroscopy in the resolved sideband regime, PRS in unresolved sideband scenarios can be interesting as well for a range of investigations of molecules in the gas phase. This includes internal state preparation, broad line absorption spectroscopy under diverse but well-controlled conditions, and single photon absorption studies of non- or weakly-fluorescing molecules.
The unresolved sideband PRS scenario appears naturally in two generic cases, when the motional sideband frequencies of the two-ion system are either smaller than or similar to: 1) the natural linewidth of the spectroscopic transition, or 2) the linewidth of the applied light source. In general, to simulate the photon recoil spectrum under such circumstances, one has to solve the very complicated master equations \cite{Stenholm1986} based on the theory presented in Sec. \ref{ResolvedPRS} but including the laser linewidth and/or the natural linewidth of the addressed transition. However in the following, we will take a simpler approach which should be valid in the limit of no remaining coherence in the interaction with the light field. The internal state evolution can then be described by excitation and de-excitation rates in accordance with Einstein's theory for light absorbers interacting with broadband (blackbody) fields \cite{Hoeppner2012}. We will further assume that the wave vectors of the absorbed and emitted photons can all be represented by the one corresponding to that of the transition center (i.e. assuming the transition linewidth to be much narrower than the transition frequency). We can then apply Eq. (\ref{scalingFactor}) using a single value of $\Vert \vec{k}_L \Vert$ (or $\Vert \vec{k}_{spon} \Vert$ for spontaneous emission) to evaluate the relative coupling between motional states.
Based on these approximations, we can now formally write up rate equations governing the dynamics of the internal and external state populations as:
\begin{widetext}
\begin{align}
\begin{split}
\dv{t} P_{\ket{\ir,\gt,\nip,\nop}} = \sum_{\sip,\sop} &- \Rabs(\nip,\nop,\sip,\sop)P_{\ket{i_r,g_t,\nip,\nop}} \\
& + [\Rstim(\nip,\nop,\sip,\sop) + R_{\text{spon}}(\nip,\nip,\sip,\sop)]P_{\ket{i_r,e_t,(\nip + \sip),(\nop + \sop)}}\\
&- (R_{H,ip} +  R_{H,op})P_{\ket{i_r,g_t,\nip,\nop}} \\
&+R_{H,ip} P_{\ket{i_r,g_t,(\nip - 1),\nop}} \\
&+R_{H,op} P_{\ket{i_r,g_t,\nip,(\nop-1)}}
\end{split}
\label{rate1}
\end{align}
\begin{align}
\begin{split}
\dv{t} P_{\ket{\ir,\et,\nip,\nop}} =  \sum_{\sip,\sop} &- [\Rstim(\nip - \sip,\nop - \sop,\sip,\sop) + \Rspon(\nip - \sip,\nop - \sop,\sip,\sop)] P_{\ket{\ir,\et,\nip,\nop}} \\
& + \Rabs(\nip - \sip,\nop - \sop,\sip,\sop) P_{\ket{i_r,g_t,(\nip - \sip),(\nop - \sop)}}\\
&- (R_{H,ip} +  R_{H,op})P_{\ket{i_r,e_t,\nip,\nop}} \\
&+R_{H,ip} P_{\ket{i_r,e_t,(\nip - 1),\nop}} \\
&+R_{H,op} P_{\ket{i_r,e_t,\nip,(\nop-1)}}
\end{split}
\label{rate2}
\end{align}
for the target ion being in the internal ground or excited state respectively. Here, $R_{\text{abs/stim}}(\nip,\nop,\sip,\sop)$ describe the rates of photon absorption and stimulated emission, and can be expressed as 
\begin{align}
\begin{split}
R_{\text{abs/stim}}(\nip,\nop,\sip,\sop)  
=&B_{\text{abs/stim}}\times\rho_{\text{eff}}(\omega_t,\omega_L)\times\lvert \xi(\eta_{ip,t},\eta_{op,t},\nip,\nop,\sip,\sop) \rvert^2 \\
=&R_{\text{abs/stim},0}(\omega_t,\omega_L)\times \lvert \xi(\eta_{ip,t},\eta_{op,t},\nip,\nop,\sip,\sop) \rvert^2
\label{RAbsStim}
\end{split}
\end{align}
\end{widetext}
with
\begin{align}
B_{\text{stim}} = \frac{\pi^2 c^3}{\hbar \omega_t^3}\Gamma_t, \quad B_{\text{abs}} = B_{\text{stim}},
\label{B}
\end{align}
$\Gamma_t$ being the spontaneous decay rate of the transition. $\rho_{\text{eff}}(\omega_t,\omega_L)$ denotes the effective spectral energy density at the transition frequency $\omega_t$ due to a laser line centered around $\omega_L$.
Generally, we can write $\rho_{\text{eff}}(\omega_t,\omega_L)$ as 
\begin{align}
\rho_{\text{eff}}(\omega_t,\omega_L) = \frac{3I_L}{c}\int_{-\infty}^{\infty}L_t(\omega',\omega_t)L_L(\omega',\omega_L) \,d\omega',
\label{N1}
\end{align}
where $I_L$ denotes the total intensity of the laser beam and $L_t(\omega,\omega_t)$ and $L_L(\omega,\omega_L)$ represent the line shape functions for the target ion transition and laser field respectively. 
The factor of 3 in this formula is introduced because we consider here, in contrast to the original scenario considered by Einstein of classical electric dipoles interacting with unpolarized and randomly propagating electromagnetic fields, a laser field with a well-defined polarization and an aligned induced electric dipole by construction. 
We assume the laser field to have a Gaussian\footnote{This laser lineshape is chosen as an example since it is common, but any lineshape can be considered.} frequency distribution with a full width at half maximum (FWHM) $\Gamma_L \equiv \sqrt{8\ln(2)}\sigma_L$ (where 2$\sigma_L$ is the full width at $1/\sqrt{e}$), and the target transition to be Lorentzian\footnote{Typical for the natural lineshape of a transition, but it could have a different shape if other processes than spontaneous emission play a significant role.} and governed by the natural decay rate $\Gamma_t$. The two line shapes can be written as
\begin{align}
L_L(\omega,\omega_L) = \frac{1}{\sqrt{2\pi}\sigma_L}e^{\frac{-(\omega-\omega_L)^2}{2\sigma_L^2}}
\end{align}
and 
\begin{align}
L_t(\omega,\omega_t) = \frac{1}{\pi}\frac{\Gamma_t/2}{(\omega-\omega_t)^2 + \Gamma_t^2/4}.
\end{align}
In Sec. \ref{Simu} we will consider two specific cases where either $\Gamma_L \ll \Gamma_t$ or $\Gamma_L \gg \Gamma_t$ in which cases $\rho_{\text{eff}}(\omega_t,\omega_L)$ reduces to
\begin{align}
\rho_{\text{eff}}^t(\omega_t,\omega_L) = \frac{3I_L}{c}L_t(\omega_L,\omega_t),
\label{rhoMg}
\end{align} 
\begin{align}
\rho_{\text{eff}}^L(\omega_t,\omega_L) = \frac{3I_L}{c}L_L(\omega_t,\omega_L),
\label{rhoMgH}
\end{align}
respectively.
It is obvious from Eq. (\ref{RAbsStim}) that we have the largest absorption and stimulated emission rates when $\omega_L = \omega_t$, which leads to, for the two cases of $\Gamma_L \ll \Gamma_t$ and $\Gamma_L \gg \Gamma_t$, the following values for $R_{\text{abs/stim},0}(\omega_t,\omega_L)$:
\begin{align}
R_{\text{abs/stim},0}^t(\omega_t,\omega_t) \equiv R_{\text{abs,0}}^{\text{res},t}&=\frac{6\pi c^2}{\hbar \omega_t^3}I_L \nonumber \\
&\equiv \Gamma_t\frac{I_L}{I_{\text{sat}}^t},
\label{R_abs_res_Mg}
\end{align}
where
\begin{equation}
I_{\text{sat}}^t \equiv \frac{\hbar\omega_t^3\Gamma_t}{6\pi c^2},
\end{equation}
and
\begin{align}
R_{\text{abs/stim},0}^L(\omega_t, \omega_t) \equiv R_{\text{abs,0}}^{\text{res},L}&= 
\frac{3\pi^{3/2} c^2}{\sqrt{2}\hbar \omega_t^3}\frac{\Gamma_t}{\sigma_L}I_L \nonumber \\
& \equiv \Gamma_t\frac{I_L}{I_{\text{sat}}^L},
\label{R_abs_res_MgH}
\end{align}
where
\begin{equation}
I_{\text{sat}}^L \equiv \frac{\sqrt{2}\hbar\omega_t^3\sigma_L}{3\pi^{3/2} c^2},
\end{equation}
respectively. Here, the intensity $I_{\text{sat}}$ is defined in both cases as the laser intensity which leads to an excitation rate $R_{\text{abs,0}}^{\text{res}}$ equal to $\Gamma_t$.

Regarding the contribution of spontaneous emission to Eq. (\ref{rate1})-(\ref{rate2}), one has to scale the rate $\Gamma_t$ with a factor accounting for the average probability to emit on a certain sideband. This factor depends on the spatial emission pattern of the specific transition. If we define $\theta \in [0,\pi]$ as the angle between the spontaneously emitted photon wave vector $\vec{k}_{spon}$ and the $z$-axis, and $\phi \in [0,2\pi]$ as the angle between the $y$-axis and the projection of $\vec{k}_{spon}$ on the $xy$-plane, we can write 
\begin{align}
\frac{\vec{k}_{spon}}{\Vert \vec{k}_{spon} \Vert} =
\begin{pmatrix}
\sin \theta \cos \phi\\
\sin \theta \sin \phi\\
\cos \theta 
\end{pmatrix}.
\label{kspon}
\end{align}
In spherical coordinates, we can then write the spontaneous emission rate as
\begin{align}
R_{\text{spon}}(\nip,\nop,\sip,\sop) = \Gamma_t D(\nip,\nop,\sip,\sop)
\label{R_spon}
\end{align}
with
\begin{align}
D(\nip,\nop,\sip,\sop) =& \nonumber \\  \int d\Omega \lvert \xi(\eta_{ip,t}&(\theta),\eta_{op,t}(\theta),n_{ip},n_{op},s_{ip},s_{op}) \rvert^2 W(\theta,\phi),
\label{D}
\end{align}
where $\vec{k}_L$ must be replaced by $\vec{k}_{spon}$ in Eq. (\ref{Lamb}) for $\eta_{ip/op,t}$. The emission pattern $W(\theta,\phi)$ is the probability that the spontaneously emitted photon propagates along the $(\theta,\phi)$ direction. 

Finally, for realistic simulation of the PRS spectra, one has to take into account heating of the motional modes due to imperfect trapping conditions, henceforth referred to at \textit{trap-induced heating}. The associated heating rates can be very mode-dependent but are typically independent of the internal state of the two ions as well as on the specific mode excitation \cite{Morigi2001}. They are introduced in Eq. (\ref{rate1})-(\ref{rate2}) by the terms containing the rates $R_{H,ip/op}$.

Although the spectroscopy laser does not exclusively address the first BSB as in the case of resolved sideband PRS (it actually addresses many sidebands simultaneously), light-ion interaction still leads to excitation of both motional modes according to Eq. (\ref{rate1})-(\ref{rate2}). The resulting motional state populations depend on the transition line profile, the laser parameters (intensity, central frequency and linewidth) as well as the time the laser light is applied.

A signal reflecting the motional mode distribution can be obtained by applying to the readout ion a sideband resolved shelving pulse of length $\tau_r$ with respect to one of the two modes (assuming no other sidebands nor the carrier transition are driven). 
The probability for the readout ion to be shelved in the $\ket{e_r}$ state is given by 
\begin{widetext}
\begin{align}
P_{\ket{e_r}}(\tau_r,\Delta_r) = \sum_{n_{ip},n_{op}} \frac{\Omega_{n_{ip},n_{op},s_{ip},s_{op}}^2}{\Delta_r^2 + \Omega_{n_{ip},n_{op},s_{ip},s_{op}}^2}\sin^2\left(\sqrt{\Delta_r^2 + \Omega_{n_{ip},n_{op},s_{ip},s_{op}}^2}\frac{\tau_r}{2}\right)P_{\ket{\gr,i_t,\nip,\nop}}
\label{shel}
\end{align}
where $\Delta_r = \omega_L - (\omega_r + s_{ip}\omega_{ip} + s_{op}\omega_{op}) = \delta_r - (s_{ip}\omega_{ip} + s_{op}\omega_{op})$ is the detuning from the sideband in question.
For $\Delta_r=0$, the probability to stay in the ground state $\ket{g_r}$ when addressing the 1st RSB of the OP mode is then
\begin{align}
\begin{split}
P_{\ket{g_r}}(\tau_r) &= 1 - \sum_{n_{ip},n_{op}} \sin^2\left(\Omega_{n_{ip},n_{op},s_{ip}=0,s_{op}=-1}\times\frac{\tau_r}{2}\right) P_{\ket{\gr,i_t,\nip,\nop}} 
\label{fluor}
\end{split}
\end{align} 
\end{widetext}
where $\Omega_{n_{ip},n_{op},s_{ip}=0,s_{op}=-1}$ indicates the Rabi angular frequency of the 1st RSB transition for a given motional state. As mentioned previously (see Eq. (\ref{proj})), the readout fluorescence signal during step iv) is directly proportional to $P_{\ket{g_r}}$. Clearly, if there is no motional excitation by the spectroscopy laser, there is also no excitation by the resolved RSB laser pulse since $P_{\ket{e_r}}=0$ for $n_{ip}=n_{op}=0$ (assuming no heating from other sources), and the readout ion fluoresces when finally driving the $\ket{g_r} - \ket{f_r}$ transition. Conversely, any excitation by the spectroscopy laser pulse leads to a reduced fluorescence signal. As is evident from Eq. (\ref{fluor}), the fluorescence reduction also depends on the duration of the resolved RSB pulse. If $\Delta_r=0$ and $\tau_{r} = \pi/\Omega_{n_{ip},n_{op},s_{ip},s_{op}}$, we drive a $\pi$-pulse from this specific motional state and on this specific sideband, meaning all population in $\ket{\gr,i_t,\nip,\nop}$ will be transferred to $\ket{\er}$.   
In the following sections we will present simulation results for the kind of PRS spectra one obtains in the unresolved sideband scenarios.

\section{Simulations of unresolved photon recoil spectra}\label{Simu}
In this section, we present simulated unresolved PRS spectra, where either the natural linewidth of the transition (Sec. \ref{naturalBroad}) or the linewidth of the laser (Sec. \ref{laserBroad}) dominates the dynamics. 
More specifically, in Sec. \ref{naturalBroad}, we consider PRS of the rather broad 3s $^2$S$_{1/2}$ - 3p $^2$P$_{3/2}$ electronic transition ($\Gamma_t/2\pi = \SI{41.8(4)}{\mega\hertz}$) of a single \Mg ion at $\lambda=279.6$ nm by a narrow laser source ($\Gamma_L/ 2\pi \lesssim \SI{1}{\mega\hertz} \ll \Gamma_t/ 2\pi$). 
Section \ref{laserBroad} is devoted to simulation of mid-infrared vibrational PRS spectra of the very narrow $\ket{\nu=0,J=1} - \ket{\nu'=1,J'=0}$ closed transition ($\Gamma_t/2\pi =2.50$ Hz)  at $\lambda=\SI{6.17}{\micro\meter}$ in the $^1\Sigma^+$ electronic ground state of \MgH by a laser source with a linewidth varying from $\Gamma_L/2\pi\sim 50$ MHz to 1 GHz.

\subsection{Simulation of unresolved PRS due to transition linewidth}\label{naturalBroad}

\begin{figure}[t!]
\includegraphics[width = 1\textwidth]{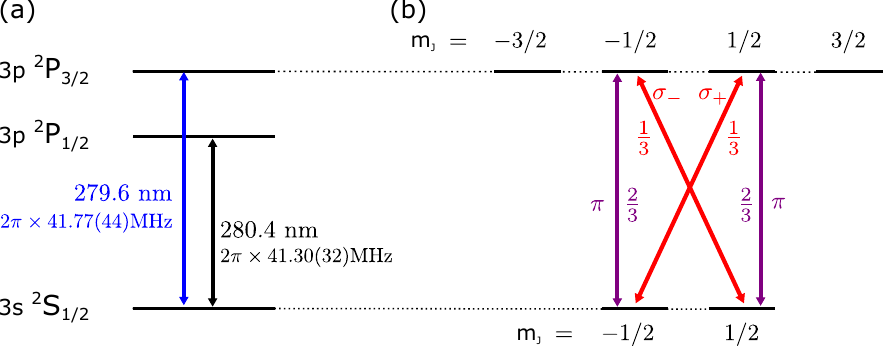}
\caption{(a) Energy levels of $^{24}$Mg$^+$. The 3s $^2$S$_{1/2}$ - 3p $^2$P$_{3/2}$ transition is the spectroscopic target transition ($\ket{g_t} - \ket{e_t}$) in our simulations and experiments. (b) The energy levels of \Mg have sub-levels which are degenerate in the presence of a magnetic field. $\pi$-transitions do not change the secondary total angular momentum $m_J$ whereas $\sigma_{\pm}$ transitions change it by $\pm 1$. The norm squared values of the Clebsch-Gordan coefficients associated with these transitions are shown.}
\label{magnesium}
\end{figure}

In this section we simulate the PRS of the rather broad 3s $^2$S$_{1/2}$ - 3p $^2$P$_{3/2}$ electronic transition ($\Gamma_t/2\pi = \SI{41.8(4)}{\mega\hertz}$) of a single $^{24}$Mg$^+$ ion at $\lambda=279.6$ nm (See Fig. \ref{magnesium}.(a)) by a laser source with such a narrow linewidth ($\Gamma_L \ll\Gamma_t$) that we disregard it. Applying linearly polarized spectroscopic laser light along a bias magnetic field axis (y-axis), only two sub-level transitions $^2$S$_{1/2}$ ($m_J=\pm 1/2$) - $^2$P$_{3/2}$ ($m_J=\pm 1/2$) can be excited, and this with equal strength (see Fig. \ref{magnesium}.(b)). Furthermore, since the spontaneous emission pattern from the $^2$P$_{3/2}$ ($m_J=\pm 1/2$) sub-states have identical effects on the motional mode excitations, the two transitions are identical from an excitation and spontaneous emission point of view, and the dynamics are equivalent to the ones of a two-level system. The effective saturation intensity is, however, 1.5 times larger than in the two-level theory presented in Sec. \ref{model} due to the norm squared of the Clebsch-Gordan coefficient being 2/3. In the simulations this has been implemented by multiplying $R_{\text{abs},0}^{\text{res}}$ from Eq. (\ref{R_abs_res_Mg}) by 2/3. 
We furthermore assume that the \Mg target ion is sympathetically cooled to the motional ground state of both modes by a directly sideband-cooled readout $^{40}$Ca$^+$ ion.

\subsubsection{Absorption and stimulated emission}

Assuming $\omega_z=2\pi\times\SI{147.9}{\kilo\hertz}$ for a single \Ca ion, we get from Eq. (\ref{modesFreqs}) $\omega_{ip}=2\pi\times\SI{162.9}{\kilo\hertz}$ and $\omega_{op}=2\pi\times\SI{300.2}{\kilo\hertz}$. Both mode angular frequencies are more than two orders of magnitude smaller than $\Gamma_t$. Hence, we are clearly in the regime where we can apply Eqs. (\ref{rate1})-(\ref{rate2}) with Eq. (\ref{rhoMg}). With $\lambda_t=\SI{279.6}{\nano\meter}$ for the target ion and the mode angular frequencies above, the LDPs for absorption and spontaneous emission are $\eta_{ip,t}=0.42$ and $\eta_{op,t}=0.51$ for the spectroscopy laser beam propagating along the $z$-axis. In the simulations to be presented below, we will, however, use the values $\eta_{ip,t}=0.30$ and $\eta_{op,t}=0.36$ in order to compare the results with recent experiments where the \Mg spectroscopy laser beam propagates at a 45$^{\circ}$ angle to the $z$-axis, reducing the LDPs by a factor $\sqrt{2}$.  While this approach gives the correct effect of the momentum recoil along the $z$-axis with respect to absorption and stimulated emission, it neglects the effect of motional excitations in the plane perpendicular to the $z$-axis. In the present unresolved sideband PRS situation, it is merely expected to lead to slightly reduced induced transition rates and a minor Doppler broadening of the target line (See details in Sec. \ref{expComparison}).   

\subsubsection{Spontaneous emission}

To simulate the effect of spontaneously emitted photons following Eq. (\ref{R_spon}-\ref{D}), we must know the emission pattern of the specific transition. To determine that, we have to take into account the particular Zeeman sub-level structure of the 3s$~^2$S$_{1/2}$ - 3p$~^2$P$_{3/2}$ electronic transition in the \Mg ion and the relative branching ratios between sub-level transitions given by the norm squared of the Clebsch-Gordan coefficients (See Fig. \ref{magnesium}.(b)).
With respect to the chosen spherical coordinate system described in connection with Eq. (\ref{kspon}), the spontaneous emission pattern is given by
\begin{align}
W(\theta,\phi) = \frac{2}{3}W_{\pi}(\theta,\phi) + \frac{1}{3}W_{\sigma}(\theta,\phi),
\end{align}
where
\begin{align}
\begin{split}
W_{\pi}(\theta,\phi) &= \frac{3}{8\pi}\sin^2(\arccos(\sin \theta \sin\phi)) \quad \text{and} \\
W_{\sigma}(\theta,\phi) &= \frac{3}{16 \pi}(1 + \sin^2\theta \sin^2\phi)
\end{split}
\end{align}
are the individual emission patterns of the two possible types of sub-level transitions, $\pi$ $(\Delta m_J = 0)$ and $\sigma$ $(\Delta m_J = \pm 1)$ respectively (see Ref. \cite{Jackson} pp. 437-439).

\subsubsection{Basis for the numerical simulations}\label{simBasis}
In the following simulations, we will assume the target and readout ions to be in the $\ket{g_t}$ and $\ket{g_r}$ states respectively and both the motional modes cooled to their ground state (i.e. $\nip=0$ and $\nop=0$). In order to perform the simulations in a reasonable time on a standard personal computer, we have limited the motional state basis to a grid corresponding to $\nip=0-19$ and $\nop=0-19$ (400 states in total). Consequently, the population will eventually be moved outside the state space for long spectroscopy pulse times $\tau_{\text{spec}}$. For the present calculations the amount of population outside the state space has been limited to 1\%, effectively limiting the duration of the dynamics evolution we can simulate. Based on Eq. (\ref{scalingFactor}), which represents the relative coupling strengths of the various sidebands, we have found that it suffices to take into account sidebands up to $s_{ip,max}= \pm5$ and $s_{op,max}=\pm6$. Based on trap-induced heating rate measurements, we use $R_{H,ip}=\SI{14(1)}{\per\second}$ and $R_{H,op}=\SI{1.7(3)}{\per\second}$.

\subsubsection{Simulation of the dynamics of motional state populations on resonance}

In the present case, we consider the limit where $\Gamma_L\ll\Gamma_t$. 
From Eq. (\ref{R_abs_res_Mg}), the saturation intensity is $I_{sat}^t=\SI{0.749}{\watt\per\centi\meter\squared}$. 
For any laser intensities $I_L$ and when trap-induced heating rates can be ignored, the evolution of the system (described by Eqs. (\ref{rate1}-\ref{rate2})) is governed by the rate of absorption and the dynamics should have exactly the same behavior when $\tau_{\text{spec}}$ is scaled by $R_{\text{abs,0}}^{\text{res,t}}$. 
To express this we define $\tau_{\text{scaled}}^t \equiv \tau_{\text{spec}} \times R_{\text{abs,0}}^{\text{res},t}$, which, for low (high) laser intensities, corresponds to the number (half the number) of absorbed photons from the laser on resonance.


\begin{figure}[t!]
\centering
\includegraphics[width = 1\textwidth]{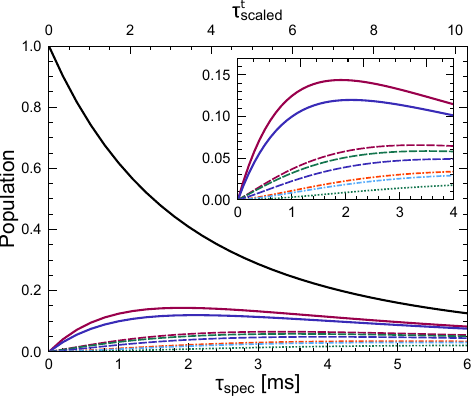}
\caption{Simulated evolution of the population in different motional states as a function of spectroscopy pulse time $\tau_{\text{spec}}$ (bottom axis) and scaled time $\tau_{\text{scaled}}^t$ (top axis) for a laser on resonance ($\omega_L = \omega_t$) and with $I_L = \SI{4.9}{\micro\watt\per\centi\meter\squared} = 6.54 \times 10^{-6} I_{\text{sat}}^t$. The following notation refers to motional states $(n_{ip},n_{op})$: (0,0) \protect\zerozero, (0,1) \protect\zeroone, (1,0) \protect\onezero, (0,2) \protect\zerotwo, (1,1) \protect\oneone, (2,0) \protect\twozero, (1,2) \protect\onetwo, (2,1) \protect\twoone, (2,2) \protect\twotwo.  Inset: zoom.}
\label{dummy1}
\end{figure}

In Fig. \ref{dummy1}, we show the temporal evolution of the population of the various motional states ($\nip$, $\nop$) from solving Eqs. (\ref{rate1}-\ref{rate2}) assuming $\omega_L=\omega_t$ and for $I_L= 6.54 \times 10^{-6} I_{sat}^t$. This laser intensity is chosen to compare with experimental results presented in sub-section \ref{expComparison}. More specifically, the figure shows the populations versus real time $\tau_{\text{spec}}$ (bottom x-axis) as well as the scaled time $\tau_{\text{scaled}}^t$ (top x-axis). 
As evident from this figure, the motional ground state population (0,0) is a monotonically decreasing function of time. For short times, the populations of all other motional states $(\nip,\nop)\neq (0,0)$ increase linearly with time as expected for rate equations. For longer times, the populations saturate and eventually decrease as a broader range of motional states is reached.

\subsubsection{Simulation of motional population spectra}

\begin{figure}[b!]
\centering
\includegraphics[width = 1\textwidth]{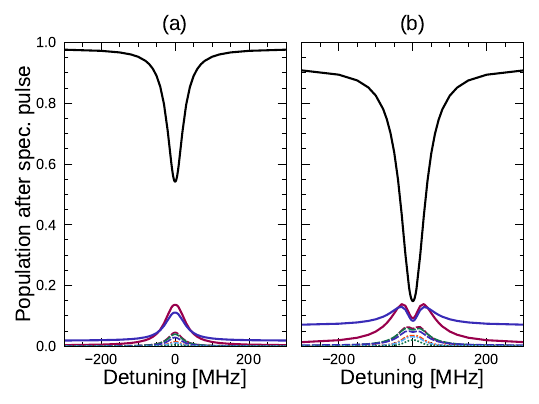}
\caption{Simulated population in different motional states after a spectroscopy pulse of length $\tau_{\text{spec}}$ = 1.3 ms ($\tau_{\text{scaled}}^t = 2.23$) \textbf{(a)} and 5.3 ms ($\tau_{\text{scaled}}^t = 9.10$) \textbf{(b)} as a function of spectroscopy laser detuning for a spectroscopy laser intensity of $I_L = \SI{4.9}{\micro\watt\per\centi\meter\squared} = 6.54 \times 10^{-6} I_{\text{sat}}^t$. The following notation refers to motional states $(n_{ip},n_{op})$: (0,0) \protect\zerozero, (0,1) \protect\zeroone, (1,0) \protect\onezero, (0,2) \protect\zerotwo, (1,1) \protect\oneone, (2,0) \protect\twozero, (1,2) \protect\onetwo, (2,1) \protect\twoone, (2,2) \protect\twotwo.}
\label{dummy2}
\end{figure}

Scanning the laser frequency $\omega_L$ across the resonance of the target ion transition for a fixed $\tau_{\text{spec}}$ leads to what we name the motional population spectra. In Fig. \ref{dummy2}, we present such spectra for $I_L= 6.54 \times 10^{-6} I_{sat}^t$ and for spectroscopy times $\tau_{\text{spec}}=\SI{1.3}{\milli\second}$ and 5.3 ms ($\tau_{\text{scaled}}^t$ = 2.23 and 9.10). For these parameters, one clearly sees the effect of population depletion of the lower excited motional states around the resonance due to the effective motional state spreading. A normal spectral response with a width reflecting the natural linewidth of the transition is observed only for the weakly populated states. For PRS one has, however, to be aware that the motional spectra presented in Fig. \ref{dummy2} will not be read out individually by the readout procedure (steps (iii) and (iv)) reported previously. This readout signal is instead an intricate combination of contributions from each of the motional population spectra, as will be discussed in the following sub-section.

\subsubsection{Simulation of readout spectra}\label{simReadout}

To simulate the PRS spectrum, one first has to solve the time dependent Schr\"{o}dinger equation Eq. (\ref{schrodinger}) in the $\ket{i_r,i_t,\nip,\nop}$ basis when addressing the readout ion on a specific RSB transition. The coupling matrix elements are given by Eq. (\ref{matrixElement}) and the initial state is the mixed state resulting from the rate equation dynamics presented in the previous sub-section. Since the fluorescence signal is sideband unresolved and hence essentially just proportional to $P_{\ket{g_r}}$, it corresponds in the simulation to projecting the final state $\ket{\psi}$ on the $\ket{g_r,i_t,n_{ip},n_{op}}$ states as described by Eq. (\ref{proj}) and (\ref{fluor}).

\begin{figure}[h!]
	\centering
	\includegraphics[width = 1\textwidth]{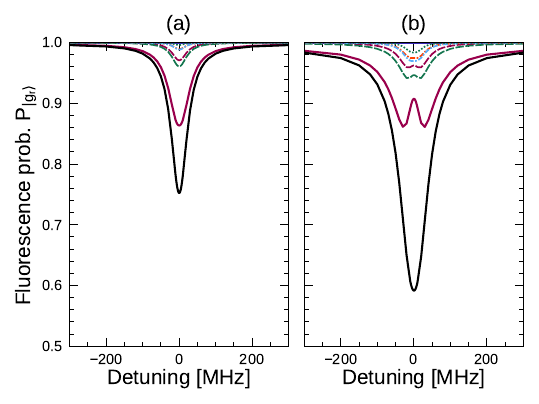}
	\caption{Simulated fluorescence probability $P_{\ket{g_r}}$, \protect\zerozero, corresponding to the detected signal for readout on the OP mode, for spectroscopy pulse lengths of $\tau_{\text{spec}}$ = 1.3 ms ($\tau_{\text{scaled}}^t = 2.23$) \textbf{(a)} and 5.3 ms ($\tau_{\text{scaled}}^t = 9.10$) \textbf{(b)} as a function of spectroscopy laser detuning for a spectroscopy laser intensity of $I_L = \SI{4.9}{\micro\watt\per\centi\meter\squared} = 6.54 \times 10^{-6} I_{\text{sat}}^t$. The other curves are the contributions from different motional states, corresponding to the terms in the sum of Eq. (\ref{fluor}). The following notation refers to motional states $(n_{ip},n_{op})$: (0,1) \protect\zeroone, (1,1) \protect\oneone,  (2,1) \protect\twoone, (0,2) \protect\zerotwo, (1,2) \protect\onetwo, (2,2) \protect\twotwo.}
\label{dummy4}
\end{figure}

In Fig. \ref{dummy4}, we show the norm of such projections corresponding to the two population spectra in Fig. \ref{dummy2} after applying a readout pulse on the \Ca ion. More specifically, this pulse corresponds to a $\pi$-pulse with respect to the (0,1) $\rightarrow$ (0,0) 1st RSB of the 4s $^2$S$_{1/2}$ - 3d $^2$D$_{5/2}$ quadrupole transition ($\tau_{r} = \pi/\Omega_{n_{ip}=0,n_{op}=1,s_{ip}=0,s_{op}=-1}$). With a transition wavelength of 729 nm and the mode frequencies given above, we get $\eta_{ip,r}=0.204$ and $\eta_{op,r}= 0.0917$.
A bit surprisingly, the fluorescence spectrum in Fig. \ref{dummy4}.(b) shows no sign of the depletion around resonance seen in the corresponding population spectrum presented in Fig \ref{dummy2}.(b). This is because the many small contributions from non-depleted higher motional states ``fill out the dip''.

\begin{figure}[b!]
\includegraphics[width = 1\textwidth]{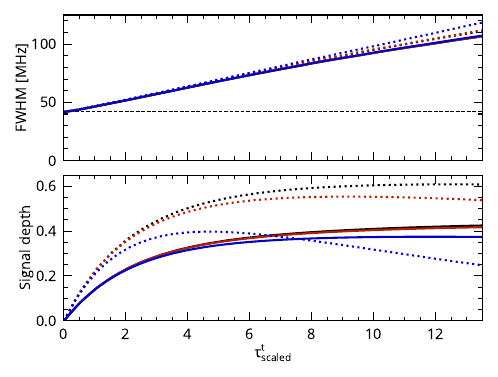}
\caption{FWHM (top) and signal depth (bottom) of simulated PRS spectra for readout on the OP mode (solid lines only) as a function of scaled time for three different spectroscopy laser intensities $I_L = \SI{1}{\micro\watt\per\centi\meter\squared} = 1.34 \times 10^{-6} I_{\text{sat}}^t$ (\protect\blueline),  $I_L = \SI{5}{\micro\watt\per\centi\meter\squared} = 6.68 \times 10^{-6} I_{\text{sat}}^t$ (\protect\redline), and $I_L = \SI{15}{\micro\watt\per\centi\meter\squared} = 2.00 \times 10^{-5} I_{\text{sat}}^t$ (\protect\blackline). The deviation between the lines is caused by trap-induced heating, which influences low intensities more, due to lower $R_{\text{abs},0}^{\text{res},t}$ values. The dashed line (top) represents the natural linewidth of the target transition. The dotted lines are the results for readout on both the OP and IP modes (see Sec. \ref{expComparison}).}
	\label{FWHM_scaled}
\end{figure}

It is clear from Fig. \ref{dummy4} that there is a spectral broadening and an increase in signal depth with $\tau_{\text{spec}}$ for a fixed $R_{\text{abs,0}}^{\text{res},t}$. However, the spectroscopic signal does not significantly change as long as $\tau_{\text{scaled}}^t$ is constant. This can be seen in Fig. \ref{FWHM_scaled} (solid lines) showing the simulated signal FWHM and depth as a function of $\tau_{\text{scaled}}^t$ for various laser intensities. The slight discrepancies between the different lines are due to trap-induced heating which, as expected, is almost negligible for the \Mg case. Indeed, trap-induced heating plays a role only for very weak laser intensities for which the light induced rate out of the motional ground state is reduced to the order of the trap-induced heating rate. 

The FWHM at scaled times $\tau_{\text{scaled}}^t \rightarrow 0$ has the expected value of the target transition linewidth and increases almost linearly with $\tau_{\text{scaled}}^t$. This is because the signal relies on the depletion of the motional ground state. Even at large detunings, there is still a small probability to excite the transition and to move population out of the motional ground state. The probability for each detuning increases with $\tau_{\text{scaled}}^t$, but eventually saturates when all the population is moved out. As $\tau_{\text{scaled}}^t$ keeps increasing, depletion of the motional ground state happens for a broader and broader frequency span around the resonance, and thus the signal width increases indefinitely.
This depletion effect also causes the observed signal depth saturation for high $\tau_{\text{scaled}}^t$. Indeed, for very long $\tau_{\text{scaled}}^t$ the population is spread out over many motional states from where the probability to not be shelved by the readout pulse is governed by Eq. (\ref{fluor}). In the limit where all the population is in many very high motional states, $P_{\ket{g_r}} \rightarrow 1/2$ (the average value of a $\sin^2(x)$ function). The signal depth can be 1 only in the resolved PRS regime, where the motional ground state population is coherently driven to one specific excited motional state.

\begin{figure}[h!]         
\begin{center}
\includegraphics[width = 1\textwidth]{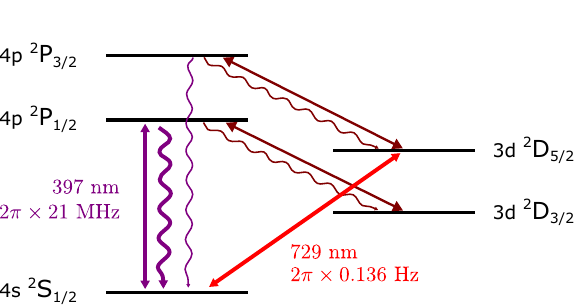}
\caption{Energy levels and electronic transitions of the \Ca readout ion relevant to laser cooling and resolved sideband spectroscopy. In the terms of Fig. \ref{expQLS}, 4s $^2$S$_{1/2}=\ket{g_r}$, 3d $^2$D$_{5/2} = \ket{e_r}$, and 4p $^2$P$_{1/2} = \ket{f_r}$.}
\label{calcium}
\end{center}
\end{figure}

\begin{figure*}[t!]          
\includegraphics[width = 1\textwidth]{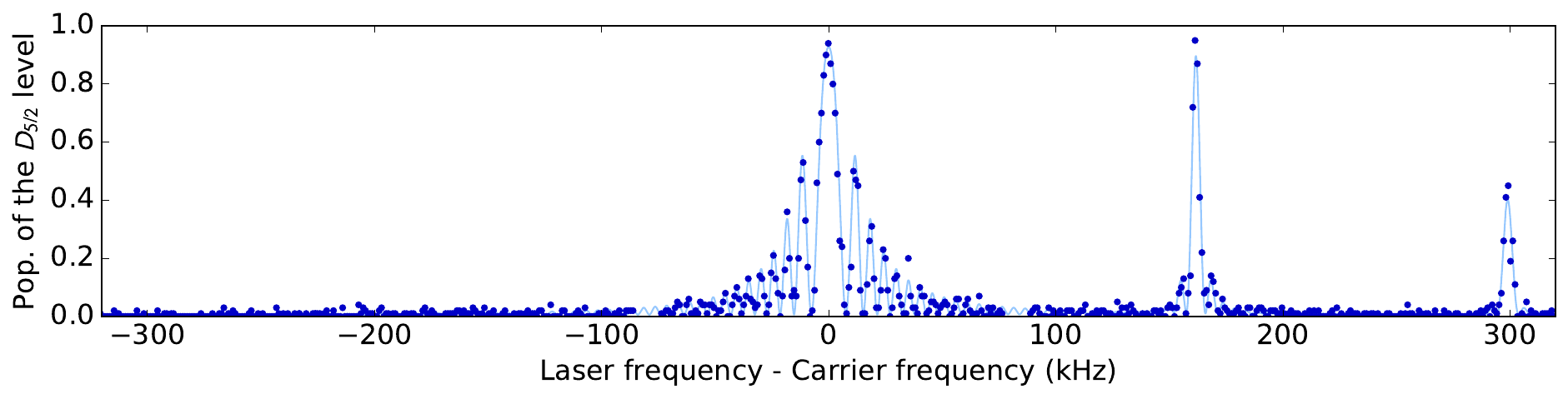}
\caption{Resolved sideband spectrum representing the probability of excitation of the 4s $^2$S$_{1/2}$ - 3p $^2$D$_{5/2}$ transition in \Ca as a function of the 729 nm laser's detuning with respect to the carrier transition. This spectrum was obtained after Doppler cooling followed by sideband cooling. The absence of visible RSBs indicates a high probability of occupation of the ground state for both motional modes. The duration and intensity of the 729 nm probe pulse were chosen to perform a $\pi$-pulse on the 1st BSB of the IP mode at $\SI{162.9}{\kilo\hertz}$. The blue line is a fit to the data points and gives mean occupation numbers of $\langle \nip\rangle=0.09(+0.18-0.09)$ and $\langle \nop\rangle=0.14(+0.24-0.14)$ for the IP and OP modes respectively.}
\label{sbcSpec}
\end{figure*}

\subsubsection{Comparison with experimentally obtained spectra}\label{expComparison} 

The experiments are initialized by sideband cooling a single \Mg and a single \Ca ion to the ground state with respect to both the IP and OP modes. Sideband cooling is achieved by addressing the 4s $^2$S$_{1/2}$ - 3d $^2$D$_{5/2}$ quadrupole transition of the \Ca ion (See Fig. \ref{calcium}). The IP and OP mode angular frequencies are $\omega_{ip}=2\pi\times\SI{162.9}{\kilo\hertz}$
and $\omega_{op}=2\pi\times\SI{300.2}{\kilo\hertz}$ respectively. A typical sideband excitation spectrum on the \Ca quadrupole transition after sideband cooling is presented in Fig. \ref{sbcSpec}. A fit to the experimental data points leads to mean occupation numbers of $\langle \nip\rangle=0.09(+0.18-0.09)$ and $\langle \nop\rangle=0.14(+0.24-0.14)$ for the IP and OP modes respectively. The laser beam exciting the \Mg ion makes a 45$^{\circ}$ angle with the $z$-axis. Its polarization is linear and aligned with the $y$-axis, along which a weak magnetic bias field (6.523(3) G) is also pointing.

\begin{figure}[h!] 
\includegraphics[width=1\textwidth]{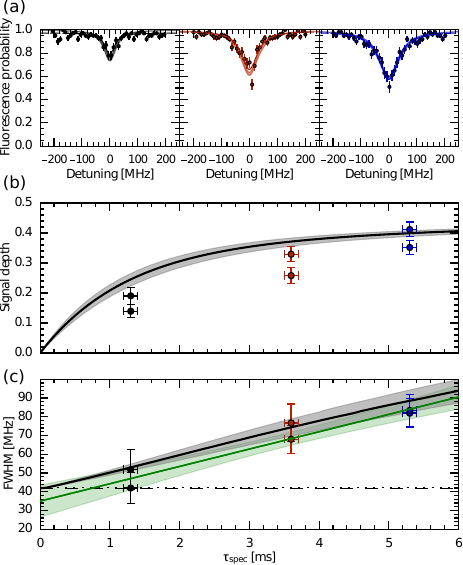} 
\caption{Unresolved sideband PRS of the 3s $^2$S$_{1/2}$ - 3p $^2$P$_{3/2}$ transition in the \Mg ion with a single co-trapped \Ca ion as the readout ion after readout on the OP mode (a) Typical experimental spectra obtained for a spectroscopy laser intensity of $I_L = \SI{4.9(7)}{\micro\watt\per\centi\meter\squared} = 6.5(9) \times 10^{-6} I_{\text{sat}}^t$ and, from left to right, spectroscopy times of $\tau_{\text{spec}}=$ 1.3(1) ms, 3.6(1) ms and 5.3(1) ms ($\tau_{\text{scaled}}^t$ = 2.2, 6.2, and 9.1). The lighter colored broader lines represent the simulated results when including the 1$\sigma$-uncertainty on the experimental value of $I_L$. The dashed lines are Lorentzian fits to the experimental data from which the signal depths and FWHM are extracted. (b) Extracted signal depth as a function of $\tau_{\text{spec}}$ together with simulated values for $I_L = \SI{4.9}{\micro\watt\per\centi\meter\squared}$ (black line) including the 1$\sigma$-uncertainty on $I_L$ (gray shaded area). (c) Extracted FWHM as a function of $\tau_{\text{spec}}$ together with the simulated values (black line). A linear fit of the experimental data is also shown in green together with the resulting one standard deviation of the fit (green shaded area). The intercept at zero spectroscopy time of 35(9) MHz matches the expected FWHM (dotted line) resulting from the natural linewidth (dashed line), the Doppler and Zeeman effects.}
\label{1pulse3graphs}
\end{figure}

Experimentally obtained PRS spectra of the 3s $^2$S$_{1/2}$ - 3p $^2$P$_{3/2}$ transition in \Mg are presented in Fig. \ref{1pulse3graphs}.(a). In the experiments we used $I_L = \SI{4.9(7)}{\micro\watt\per\centi\meter\squared} = 6.5(9) \times 10^{-6} I_{\text{sat}}^t$ and, from left to right, spectroscopy excitation times of $\tau_{\text{spec}}$ = 1.3(1), 3.6(1) and 5.3(1) ms. 
All three experimental spectra have been centered at a detuning of 0 MHz, while the wavelength meter readings gave offsets of 12(4) MHz, 11(2) MHz and 18(3) MHz respectively, compared to the most precisely measured value of the transition \cite{Herrmann2009}. These discrepancies are however all within the accuracy of the wavelength meter (HighFinesse \AA ngstom WS-U). Figs. \ref{1pulse3graphs}.(b) and \ref{1pulse3graphs}.(c) show, together with simulation results, the evolution of the measured spectral depth and FWHM as a function of spectroscopy time $\tau_{\text{spec}}$. There is a fairly good agreement between experiments and simulations within the error bars. 
The black dashed line in Fig. \ref{1pulse3graphs}.(c) represents the natural FWHM of the transition ($\Gamma_t/2\pi = \SI{41.8(4)}{\mega\hertz}$). The black dotted line represents the effective FWHM which is slightly broader due to two effects. 
Firstly, the higher temperature ($\sim$ 0.75 mK) of the ions perpendicular to the $z$-axis leads to a Doppler broadening in the direction of the applied spectroscopic laser beam. This broadening is about $\Delta\omega_{Dopp}\sim 2\pi\times\SI{3}{\mega\hertz}$ and is common to each of the two ($m_J=\pm 1/2$) - ($m_J=\pm 1/2$) sub-level transitions. Secondly, the applied weak magnetic field gives rise to a differential Zeeman shift of the two sub-level transitions of $\Delta\omega_{Zee}= 2\pi\times\SI{6.1}{\mega\hertz}$. By simulating the effective line profile of the transition taking both effects into account, we find an effective FWHM of about 42.5 MHz. A linear fit of the experimental data gives a FWHM at zero spectroscopy time of 35(9) MHz (green line Fig. \ref{1pulse3graphs}.(c)). Although not a precision measurement, this result matches the effective FWHM of the transition well.


\begin{figure} 
\includegraphics[width=1\textwidth]{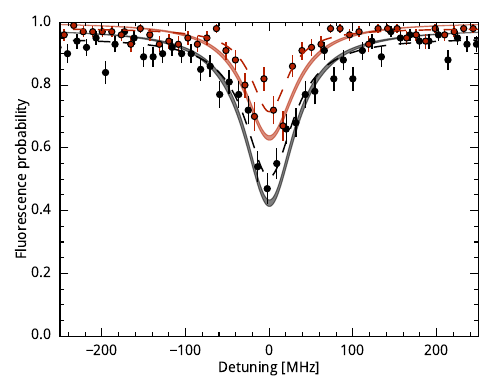}
\caption{Comparison between the photon recoil spectra of the 3s $^2$S$_{1/2}$ - 3p $^2$P$_{3/2}$ transition in \Mg obtained adressing either only the OP mode (red) or both the OP and IP mode (black) before readout. The lighter colored broader lines represent the simulated results when including the $1\sigma$-uncertainty of the experimental value of the spectroscopy laser light intensity $I_L$, while the dashed lines are Lorentzian fits to the experimental data. A $\sim50$\% increase in signal depth is clearly seen. $I_L = \SI{10(1)}{\micro\watt\per\centi\meter\squared} = 1.3 \times 10^{-5} I_{sat}^t$ and $\tau_{\text{spec}} = $ 1.6(1) ms ($\tau_{\text{scaled}}^t = 5.6$). Note that $I_L$ was increased by a factor two compared to the spectra in Fig. \ref{dummy4} and \ref{1pulse3graphs}, to not be limited by the faster heating rate of the IP mode.}
\label{spectro2pulses}
\end{figure}

Since the IP and OP modes become similarly excited during the spectroscopy pulse, as is clear from Fig. \ref{dummy1}, the spectroscopic signal depth might actually be increased significantly by addressing not only the OP mode but also the IP mode during the readout phase. This scenario can be realized by consecutively applying $\pi$-pulses on the (0,1) $\rightarrow$ (0,0) and (1,0) $\rightarrow$ (0,0) sideband transitions ($\tau_{r}^{\text{1st}} = \pi/\Omega_{n_{ip}=0,n_{op}=1,s_{ip}=0,s_{op}=-1}$ and $\tau_{r}^{\text{2nd}} = \pi/\Omega_{n_{ip}=1,n_{op}=0,s_{ip}=-1,s_{op}=0}$), with respect to two different upper sub-levels of the $\ket{g_r} - \ket{e_r}$ transition. The  result using two such readout pulses is shown in Fig. \ref{FWHM_scaled} (dotted lines). We see that the FWHM of the expected signal does not differ much from the one-pulse readout result (solid lines), but that the signal depth is significantly increased as long as the spectroscopy laser intensity is large enough for the laser induced motional excitation to be faster than the IP heating rate. For example, for the smallest shown intensity (blue), the signal depth of the two-pulse readout is for $\tau_{\text{scaled}}^t>8$ smaller than for the one-pulse readout and gets worse for longer times.
In the case of \Ca being the readout ion, one can in principle sequentially address up to four sub-level transitions (e.g. $m_J=-1/2$ $\rightarrow$ $m_{J'}=-5/2,-3/2,1/2,3/2$) by different orders of RSB pulses for both motional modes, and in this way increase the spectroscopy signal even further. However, 
the more RSB pulses the longer the readout phase will become and hence trap-induced heating will eventually limit this strategy.
In Fig. \ref{spectro2pulses}, we present experimental PRS spectra when addressing either only the (0,1) $\rightarrow$ (0,0) sideband or both the (0,1) $\rightarrow$ (0,0) and (1,0) $\rightarrow$ (0,0) sidebands. In this particular case, one clearly sees the gain in signal depth in the two $\pi$-pulse scheme.
In the experiment, we used $I_L= \SI{10(1)}{\micro\watt\per\centi\meter\squared} \approx 1.3 \times 10^{-5} I_{sat}^t$ with a spectroscopy excitation time of $\tau_{\text{spec}}$ = 1.6(1) ms ($\tau_{\text{scaled}}^t = 5.6$). Note that the intensity of the spectroscopic pulse was increased by a factor two compared to the spectra in Fig. \ref{1pulse3graphs}, in order to not be limited by the faster heating rate of the IP mode as compared to the OP mode.
Both experimental spectra have been centered at a detuning of 0 MHz, while the wavelength meter readings gave offsets of 16(3) MHz and 20(3) MHz, respectively. These discrepancies are once again both within the inaccuracy of the wavelength meter.

\subsection{Simulation of unresolved PRS due to laser linewidth}\label{laserBroad}

\begin{figure*} 
\includegraphics[width = 1\textwidth]{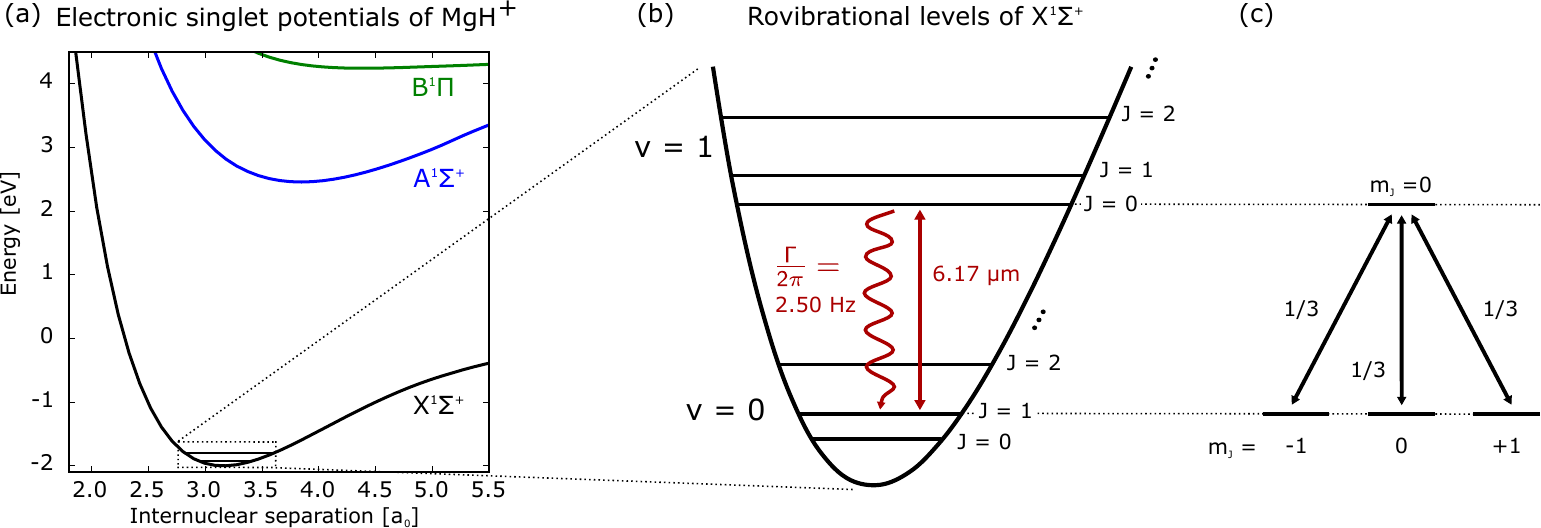}
\caption{(a) The electronic singlet potentials of MgH$^+$. Only the electronic ground state $X^1\Sigma^+$ is considered in the simulation. (b) Rovibrational structure of the electronic ground state $X^1\Sigma^+$ showing the closed transition of interest $\ket{v=0,J=1} - \ket{v'=1,J'=0}$. The decay rate of $\Gamma_t =2\pi\times 2.50$ Hz is the slowest timescale of our experiment and spontaneous emission is negligible. (c) Sub-levels of the $\ket{v=0,J=1}$ and $\ket{v'=1,J'=0}$ states showing the norm squared of the Clebsch-Gordan coefficients.}
\label{vibrationLevels}
\end{figure*}

In this section we present simulation results of mid-infrared vibrational PRS of the very narrow $\ket{v=0,J=1}$ - $\ket{v'=1,J'=0}$ transition ($\Gamma_t/2\pi =2.50$ Hz)  at $\SI{6.17}{\micro\meter}$ in the $^1\Sigma^+$ electronic ground state of \MgH (See Fig. \ref{vibrationLevels}) by laser sources with varying linewidths from $\Gamma_L\sim2\pi\times\SI{50}{\mega\hertz}$ to 1 GHz. As for the \Mg ion, the considered internal state structure does not really constitute a two-level system but a four-level one as shown in Fig. \ref{vibrationLevels}.(c). However, choosing for example a linearly polarized light source, absorption can only happen from a single rotational sub-state, the $m_J=0$ state. Similarly, stimulated emission can only happen back to the same sub-state and an effective two-level scheme is established with respect to interactions with the light field. However, as in the case of the \Mg ion, the saturation intensity has to be scaled by the norm squared of the relevant Clebsch-Gordan coefficient. This is implemented by dividing $R_{\text{abs},0}^{\text{res},L}$ by 3 for both absorption and stimulated emission. The present level scheme has the further complication that population in the $m_J=\pm 1$ sub-states does not interact with the light field. Hence, if the molecular ion was originally in one of these sub-states (or if one of these sub-states were populated through spontaneous emission), it would not contribute (anymore) to the PRS signal. To avoid this effect, one can apply a magnetic field not aligned with the polarization axis of the light source, which would lead to Larmor precession of populations between the three sub-states of the $\ket{v=0,J=1}$ level. The situation becomes particularly simple when the Larmor frequency $\omega_{Larmor}\gg R_{abs}$. In this case, one can assume the total lower state population at any instance to be equally distributed between the three $m_J=0,\pm 1$ sub-states, i.e. one third in each. Therefore we can reestablish an effective two-level scenario, but with the absorption rate (and not the stimulated emission rate) divided by another factor of 3. This leads to a saturation intensity 9 times larger than the one presented in Eq. (\ref{R_abs_res_MgH}). In the following simulations, we assume this picture to be true and also disregard the hyperfine splitting of the involved rotational levels. Indeed, the splitting is typically $\sim\SI{10}{\kilo\hertz}$, which is much smaller than the linewidth of the light source. All hyperfine components are thus addressed and are hence of no importance for the PRS signal.

\subsubsection{Absorption and stimulated emission}

We assume again $\omega_z=2\pi\times\SI{147.9}{\kilo\hertz}$ for a single \Ca ion, which leads to $\omega_{ip}=2\pi\times\SI{162.0}{\kilo\hertz}$ and $\omega_{op}=2\pi\times\SI{295.7}{\kilo\hertz}$ (only slightly different than for \Mgu). Both modes angular frequencies are at least about two orders of magnitude smaller than $\Gamma_L$. We are thus clearly in the regime where we can apply Eqs. (\ref{rate1})-(\ref{rate2}) with Eq. (\ref{rhoMgH}). With $\lambda_t=\SI{6.17}{\micro\meter}$ for the target ion and the above mode angular frequencies, the LDPs are $\eta_{ip,t}=0.0192$ and $\eta_{op,t}=0.0224$ for the spectroscopy laser beam propagating along the $z$-axis. In the simulations to be presented below, we use the values $\eta_{ip,t}=0.0136$ and $\eta_{op,t}=0.0159$ in order to eventually compare these results with experiments that we are currently setting up. Here, the laser beam makes a 45$^{\circ}$ angle with the $z$-axis, and hence the LDPs are reduced by a factor $\sqrt{2}$. For the same reasons as stated in Sec. \ref{naturalBroad}, the simulation results presented here should still be representative of the expected experimental signals.

\subsubsection{Spontaneous emission}

To simulate the effect of spontaneously emitted photons, according to Eq. (\ref{D}) we have to take into account the particular emission pattern of the $\ket{\nu=0,J=1} - \ket{\nu'=1,J'=0}$  transition (See Fig. \ref{vibrationLevels}(c)). From the upper level, the spontaneous emission pattern is completely isotropic (i.e. angle independent), since it can decay to all lower sub-levels. The general form is thus 
\begin{align}
W(\theta,\phi) = \frac{1}{4\pi}.
\end{align}

\subsubsection{Basis for the numerical simulation}

The basis for the simulations is essentially the same as in Sec. \ref{simBasis}.
However, based on Eq. (\ref{scalingFactor}), which represents the relative coupling strengths of the various sidebands, we have found that, in this case, it suffices to only take into account sidebands up to $s_{ip,max}= \pm1$ and $s_{op,max}=\pm1$.
Because the mass ratios of the \Ca - \Mg and \Ca - \MgH systems are almost the same, we do not expect the mode coupling and thereby the trap-induced heating rates to be significantly different \cite{Morigi2001}. We thus use the same values of $R_{H,ip}=\SI{14(1)}{\per\second}$ and $R_{H,op}=\SI{1.7(3)}{\per\second}$, when modeling the \Ca - \MgH system. 

\subsubsection{Simulation of the dynamics of motional state populations on resonance}

\begin{figure}[b!]        
\includegraphics[width = 1\textwidth]{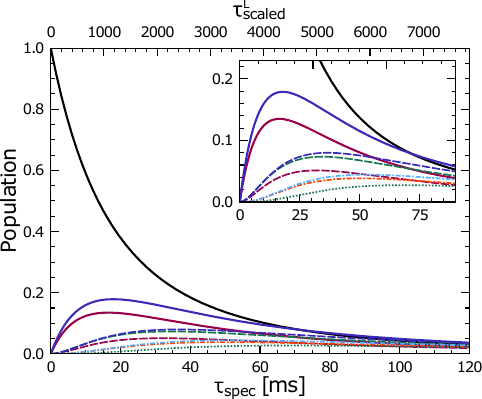}
\caption{Simulated evolution of the population in different motional states as a function of spectroscopy time $\tau_{\text{spec}}$ (bottom axis) and scaled time $\tau_{\text{scaled}}^L$ (top axis) on resonance. The spectroscopy laser FWHM is $\Gamma_L = 2\pi \times 250$ MHz and intensity $I_L = \SI{7}{\watt\per\centi\meter\squared} =  4 \times 10^{3} I_{\text{sat}}^L$. The following notation refers to motional states $(n_{ip},n_{op})$: (0,0) \protect\zerozero, (1,0) \protect\onezero, (0,1) \protect\zeroone, (2,0) \protect\twozero, (0,2) \protect\zerotwo, (1,1) \protect\oneone, (2,1) \protect\twoone, (1,2) \protect\onetwo, (2,2) \protect\twotwo. Inset: zoom.}
\label{MgHPopScaled}
\end{figure}

In contrast to the simulations discussed in Sec. \ref{naturalBroad}, here $\Gamma_L\gg\Gamma_t$ and according to Eq. (\ref{R_abs_res_MgH}) the saturation intensity $I_{sat}^L$ depends on the laser linewidth. 
Equivalently to the \Mg case, we introduce the scaled time $\tau_{\text{scaled}}^L \equiv \tau_{\text{spec}} \times R_{\text{abs,0}}^{\text{res},L}$. In Fig. \ref{MgHPopScaled}, we show the evolution of the populations of the various motional states ($\nip$, $\nop$) on resonance ($\omega_L=\omega_t$) as a function of real time $\tau_{\text{spec}}$ (bottom x-axis) and scaled time $\tau_{\text{scaled}}^L$ (top x-axis). Here $\Gamma_L/2\pi= 250$ MHz and $I_L= 7$ W cm$^{-2}$ $=4 \times 10^{3} I_{sat}^L$ with $I_{sat}^L = 1.7$ mW cm$^{-2}$ which correspond to typical parameters of our experiment.
As in Fig. \ref{dummy1}, the motional ground state population (0,0) is a monotonically decreasing function of time. For short times the populations of all other motional states $(\nip,\nop)\neq (0,0)$ increase linearly with time, while for longer times, the rate of increasing population saturates and eventually decreases as a broader range of motional states is reached. The main reason for the difference in the dynamics of the excited motional state populations at a given scaled time, as compared to the \Mg case, is the difference in motional excitations per scattered photon due to the difference in LPDs (see detailed discussion in Sec. \ref{Discussion}).

\subsubsection{Simulation of motional population spectra}
\label{sec:MgH_mot_spec}

\begin{figure}
\includegraphics[width = 1\textwidth]{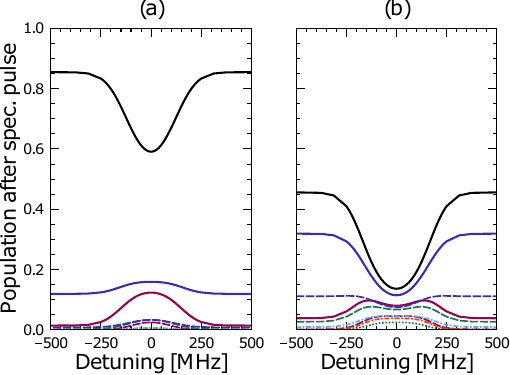}
\caption{Simulated population in different motional states as a function of spectroscopy laser detuning, after spectroscopy times $\tau_{\text{spec}}$ of 10 ms ($\tau_{\text{scaled}}^L$ = 655) \textbf{(a)} and 50 ms ($\tau_{\text{scaled}}^L$ = 3277) \textbf{(b)}. The FWHM of the spectroscopy laser is $\Gamma_L = 2\pi \times 250$ MHz and the intensity is $I_L = \SI{7}{\watt\per\centi\meter\squared} =  4 \times 10^{3} I_{\text{sat}}^L$. The following notation refers to motional states $(n_{ip},n_{op})$: (0,0) \protect\zerozero, (1,0) \protect\onezero, (0,1) \protect\zeroone, (2,0) \protect\twozero, (0,2) \protect\zerotwo, (1,1) \protect\oneone, (2,1) \protect\twoone, (1,2) \protect\onetwo, (2,2) \protect\twotwo.}
\label{mot_pop_MgH}
\end{figure}

By scanning the laser frequency $\omega_L$ across the resonance of the $\ket{v=0,J=1}$ - $\ket{v'=1,J'=0}$ transition, we obtain the motional population spectra. In Fig. \ref{mot_pop_MgH}, we present two such spectra for $\Gamma_L/2\pi = 250$ MHz, $I_L = \SI{7}{\watt\per\centi\meter\squared}=4 \times 10^{3} I_{sat}^L$ and for spectroscopy times $\tau_{\text{spec}}=\SI{10}{\milli\second}$ and $\SI{50}{\milli\second}$ ($\tau_{\text{scaled}}^L = $ 655 and 3277). 
For these parameters one clearly sees the effect of motional state depletion, but the higher weakly populated states now have a spectral response reflecting the Gaussian line shape of the laser. Another difference from the \Mg case is the increased background stemming from trap-induced heating, effectively decreasing the signal depth after readout. As evident from the solid blue and red curves, the much higher trap-induced heating rate of the IP mode (blue curve) results in a much smaller spectral depth. This clearly shows the advantage of performing readout on the OP mode instead of the IP mode, and that the two-pulse technique presented in Sec. \ref{expComparison} where both modes are addressed is not beneficial for the specific parameters. However, it can be for larger $R_{\text{abs,0}}^{\text{res},L}$. Yet, for a large enough laser beam waist of $\sim \SI{200}{\micro\meter}$ for easy alignment on $^{24}$MgH$^+$, we are experimentally limited by the available laser power to $I_L = \SI{20}{\watt\per\centi\meter\squared}$ and since we want to apply a broad laser linewidth to search for the target transition, we conclude that detection on the OP mode only is the best initial strategy to obtain a signal for the present example case. Once the target transition has been determined better such that $\Gamma_L$ can be decreased, the two-pulse technique will quickly become beneficial. 

\subsubsection{Simulation of readout spectra}

\begin{figure}[h]
\includegraphics[width = 1\textwidth]{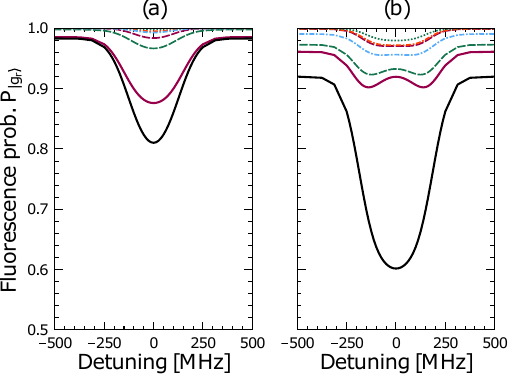}
\caption{Simulated fluorescence probability $P_{\ket{g_r}}$, \protect\zerozero, corresponding to the expected signal for readout on the OP mode, for spectroscopy pulse durations $\tau_{\text{spec}}$ of 10 ms ($\tau_{\text{scaled}}^L$ = 655) \textbf{(a)} and 50 ms ($\tau_{\text{scaled}}^L$ = 3277) \textbf{(b)} as a function of spectroscopy laser detuning, for a FWHM of $\Gamma_L = 2\pi \times 250$ MHz and intensity of $I_L = \SI{7}{\watt\per\centi\meter\squared} =  4 \times 10^{3} I_{\text{sat}}^L$ on resonance. Contributions from various motional states $(n_{ip},n_{op})$ corresponding to the terms in the sum of Eq. (\ref{fluor}) are also shown: (0,1) \protect\zeroone, (1,1) \protect\oneone,  (2,1) \protect\twoone, (0,2) \protect\zerotwo, (1,2) \protect\onetwo, (2,2) \protect\twotwo.}
\label{signal_MgH}
\end{figure}

To simulate the PRS spectrum, we follow the same procedure as in Sec. \ref{simReadout}. 
In Fig. \ref{signal_MgH}, we show readout spectra corresponding to the two population spectra in Fig. \ref{mot_pop_MgH} after first having applied the readout pulse. The latter corresponds to a $\pi$-pulse with respect to the (0,1) $\rightarrow$ (0,0) 1st RSB of the \Ca quadrupole transition. With a transition wavelength of 729 nm, and the mode frequencies given above, we get $\eta_{ip,r}=0.203$ and $\eta_{op,r}= 0.0949$.
As for the \Mg case the fluorescence spectrum in Fig. \ref{signal_MgH}.(b) shows no sign of the depletion around resonance that could be seen in the corresponding population spectra presented in Fig \ref{mot_pop_MgH}.(b).

\begin{figure}[h]
\includegraphics[width = 1\textwidth]{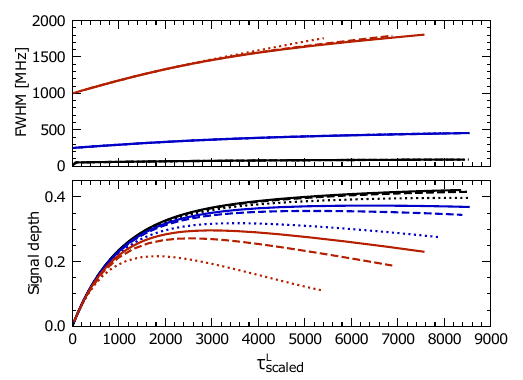}
\caption{FWHM (top) and signal depth (bottom) of simulated fluorescence probability spectra for readout on the OP mode as a function of scaled time for three different spectroscopy laser FWHM $\Gamma_L/2\pi$ of 50 MHz (\protect\blackline), 250 MHz (\protect\blueline), and 1 GHz (\protect\redline) and three different intensities: $I_L = \SI{20}{\watt\per\centi\meter\squared}$ (solid), $I_L = \SI{14}{\watt\per\centi\meter\squared}$ (dashed), and $I_L = \SI{7}{\watt\per\centi\meter\squared}$ (dotted). The deviation in signal depth for the different intensities is caused by trap-induced heating, which influences low intensities and broad spectroscopy laser linewidths more due to slower spectroscopy laser heating.}
	\label{FWHM_depth_MgH}
\end{figure}

It is clear from Fig. \ref{signal_MgH} that there is a spectral broadening and an increase in signal depth with $\tau_{\text{spec}}$ for a fixed $R_{\text{abs,0}}^{\text{res},L}$. However, the spectroscopic signal does not significantly change as long as $\tau_{\text{scaled}}^L$ is constant. This can be seen in Fig. \ref{FWHM_depth_MgH} showing the simulated signal FWHM and depth as a function of $\tau_{\text{scaled}}^L$ for various laser intensities and linewidths $\Gamma_L$. 
The discrepancies between the different lines are due to trap-induced heating. The FWHM at scaled times $\tau_{\text{scaled}}^L\rightarrow 0$ has the expected value of the laser linewidth, and then increases almost linearly with $\tau_{\text{scaled}}^L$.

\subsubsection{Detection issues due to BBR coupling}

In all previous sub-sections of Sec. \ref{laserBroad}, we considered the \MgH ion to be initially prepared in one of the two states forming the closed target transition $\ket{v=0,J=1}$ or $\ket{v'=1,J'=0}$. These states are, however, generally coupled to the rest of the rovibrational manifolds by black-body radiation (BBR) \cite{Rot2010}. Owing to the large vibrational constant of the $^1\Sigma^+$ electronic ground state at room temperature, the molecular ion is in the vibrational ground state $v=0$ with $>99.9$ \% probability and one can assume the average population of the $\ket{v=0,J=1}$ state to be equal to the thermal equilibrium value $P_\text{J=1}= 8.5$\% (at $\SI{293}{\kelvin}$).
One can further assume that if the molecular ion is initially in the $\ket{v=0,J=1}$ state, it will stay in the $\{\ket{v=0,J=1}$, $\ket{v'=1,J'=0}\}$ sub-space during the spectroscopy pulse, since $\tau_{\text{spec}}$ is much shorter than the timescale ($\sim$1 s) required for BBR or spontaneous emission to change the internal state.

\begin{figure}[h]
	\includegraphics[width = 1\textwidth]{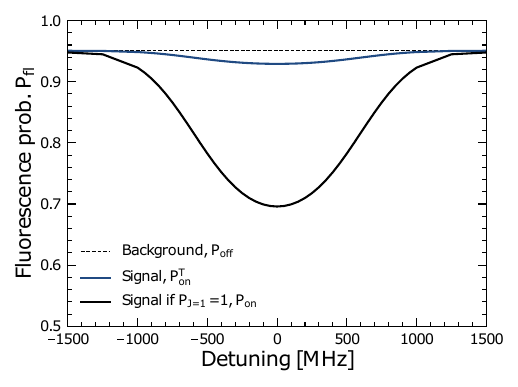}
	\caption{Comparison between the expected signal $P_\text{on}^\text{T}$ when the probability to find the molecule in the lower level of the target state $\ket{v=0,J=1}$ is $P_\text{J=1}=8.5$\%, and the expected signal $P_\text{on}$ if $P_\text{J=1}=100$\%, along with the background from trap-induced heating $P_\text{off}$ for $I_L = \SI{20}{\watt\per\centi\meter\squared}$, $\Gamma_L/(2\pi) =$ 1 GHz, and $\tau_{\text{spec}} = 30$ ms.}
	\label{signal_compare}
\end{figure}

With a probability to be in the internal target state $\ket{v=0,J=1}$ of $P_\text{J=1} = 8.5$\%, the effective signal depth is actually much smaller than the signal depth predicted in the previous sub-sections and this affects the total measurement time needed to obtain a good signal to noise ratio. 
In an experiment, the PRS sequence (i)-(iv) is repeated $N$ times for each frequency point (i.e for each value of the detuning $\omega_L-\omega_t$). The total measurement time for a given frequency point is thus $T_\text{freq}=N\tau_\text{cycle}$, where $\tau_{\text{cycle}}$ (sometimes called the experiment cycle time) is the time of the PRS sequence (i)-(iv).  
Averaging over these $N$ realizations gives an estimate of the fluorescence probability of the \Ca ion which reflects the excitation probability of the \MgH ion. The probability of fluorescence when the spectroscopy laser is far from resonance is denoted $P_\text{off}$ (background). It depends solely on trap-induced heating and imperfect ground state cooling. The probability of fluorescence close to resonance is denoted $P_\text{on}$ (signal). We have $P_\text{on}\leq P_\text{off}$, with strict inequality if the target transition is driven.
The reduced population in the $\ket{v=0,J=1}$ state due to coupling to BBR and spontaneous emission can be taken into account by scaling the simulated signal depth $P_\text{off}-P_\text{on}$ by $P_\text{J=1}$, such that the effective signal fluorescence probability $P_\text{on}^\text{T}$ becomes
\begin{align}
\begin{split}
P_\text{on}^\text{T} &= P_\text{off} - [P_\text{off} - P_\text{on}]P_\text{J=1}\\
&= (1-P_\text{J=1})P_\text{off} + P_\text{J=1}P_\text{on}.
\end{split}
\label{P_on_T}
\end{align}
In Fig. \ref{signal_compare} we compare $P_\text{on}$ and $P_\text{on}^\text{T}$ for $I_L=$ $\SI{20}{\watt\per\centi\meter\squared}$, $\Gamma_L/(2\pi) =$ 1 GHz, $\tau_{\text{spec}} = 30$ ms, and $T=\SI{293}{\kelvin}$. 

A relevant criteria for distinguishing the signal ($P_\text{on}^\text{T}$) from the background ($P_\text{off}$) is
\begin{align}
P_\text{off} - P_\text{on}^\text{T} > C(\sigma_\text{off} + \sigma_\text{on}^\text{T})
\label{criteria}
\end{align}
where $C$ is the confidence level expressed in number of standard deviations, and $\sigma_\text{off}$ and $\sigma_\text{on}^\text{T}$ are the uncertainties on the measured values of $P_\text{off}$ and $P_\text{on}^\text{T}$, respectively. Assuming Gaussian uncorrelated statistics of the parameters in Eq. (\ref{P_on_T}) we can estimate the uncertainty of the thermal state signal as
\begin{align}
\begin{split}
\sigma_\text{on}^{T} &= [ (1-P_\text{J=1})^2\sigma_\text{off}^2 + (P_\text{J=1})^2\sigma_\text{on}^2 \\
&+ (P_\text{on} - P_\text{off})^2 (\sigma_\text{J=1})^2 ]^{1/2}
\end{split}
\label{sigma_on_thermal}
\end{align}
where $\sigma_\text{on}$ ($\sigma_\text{off}$) is the binomial uncertainty on $P_\text{on}$ ($P_\text{off}$), which depends on the number of realizations $N$ as $\sigma_\text{on/off} = \sqrt{P_\text{on/off}(1-P_\text{on/off})/N}$. $\sigma_\text{J=1}$ is the uncertainty on $P_\text{J=1}$ which depends on the total measurement time $T_\text{freq}$ as $\alpha/\sqrt{T_\text{freq}}$ where $\alpha$ is a factor related to the lifetime of the $J=1$ state, $\tau_\text{J=1}$. A Monte-Carlo simulation of the internal state dynamics taking into account all coupling rates between the different rotational states (resulting from BBR and spontaneous emission) gives $\alpha \sim 0.8$ as long as $\tau_{\text{cycle}}$ is small compared to $\tau_\text{J=1} \sim 3$ s. 
Of the three terms in Eq. (\ref{sigma_on_thermal}), the latter including $\sigma_\text{J=1}$ is the dominating one for typical values of $P_\text{off} = 0.95$, $P_\text{on} = 0.60$ and $\tau_{\text{cycle}} = 0.1$ s. Therefore, in this case, one does not gain from reducing the experiment cycle time. Instead one may want to increase it (keeping $\tau_{\text{cycle}}\ll\tau_\text{J=1}$) to perform control experiment cycles (to verify state readout, ground state cooling, ...) or to measure critical parameters (laser powers, ...).      

\begin{figure}[h!]
	\includegraphics[width = 1\textwidth]{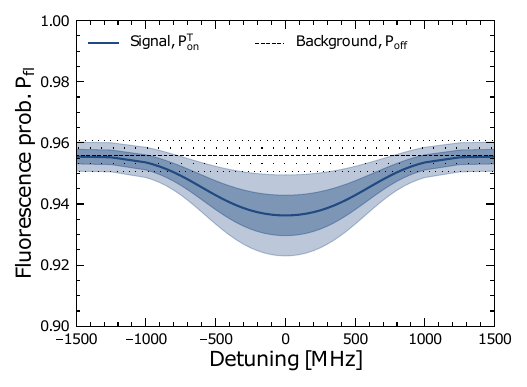}
	\caption{Simulation of the expected signal with $P_\text{J=1}=8.5$\% and background for  $I_L = \SI{20}{\watt\per\centi\meter\squared}$, $\Gamma_L/(2\pi) =$ 1 GHz, and $\tau_{\text{spec}} = 30$ ms. 
		For the signal the shaded areas represent the expected 1 (dark blue) and 2 (light blue) standard deviation measurement uncertainties, and equivalently for the background (sparsely/densely spaced dotted lines).
		The uncertainties are calculated for $N$ = 6337 experimental cycles, which is the minimum number needed to fulfill Eq. (\ref{criteria}) with $C=2$ for the given parameters. With $\tau_{\text{cycle}} = 140$ ms this results in a total time per frequency point of $\sim$15 min. 
	}
	\label{distinction}
\end{figure}

For each frequency point, the total measurement time needed to distinguish the signal from background $T_\text{freq}^\text{dist}$ can be found by solving Eq. (\ref{criteria}) for $T_\text{freq}$, using that $N =  T_\text{freq} /\tau_{\text{cycle}}$ with $\tau_{\text{cycle}} = \tau_{\text{spec}} + \tau_{\text{cd}}$, where $\tau_{\text{cd}} = \tau_{\text{(i)}}+\tau_{\text{(iii)}}+\tau_{\text{(iv)}}$ is the total time for steps (i), (iii) and (iv) of the PRS sequence (and $\tau_{\text{spec}} = \tau_{\text{(ii)}}$). Since $\tau_{\text{cd}} \sim 100$ ms in our experiments and $\tau_{\text{spec}} \sim 10$ ms, $\tau_{\text{cycle}} \ll$ $\tau_{J=1}$ and we are thus not limited by our experiment cycle time. 
Because $\tau_{\text{cycle}}(\tau_{\text{spec}})$, $P_\text{off}(\tau_{\text{spec}})$ and $P_\text{on}(\tau_{\text{spec}},\Gamma_L, I_L)$ ; $T_\text{freq}^\text{dist}$ depends only on three independent variables: $\tau_{\text{spec}}$, $\Gamma_L$ and $I_L$. 
To minimize $T_\text{freq}^\text{dist}$, $I_L$ should be maximized and $\Gamma_L$ should be minimized. Considering a reasonable laser beam waist of $\sim \SI{200}{\micro\meter}$ for easy alignment on the molecular ion, the maximum value of $I_L$ is limited by the available laser power to $\sim\SI{20}{\watt\per\centi\meter\squared}$. 
The value of $\Gamma_L$ must remain comparable to the uncertainty on the line position so that the laser frequency can be confidently tuned close to resonance. With a $1\sigma$ uncertainty on the $\ket{v=0,J=1} - \ket{v'=1,J'=0}$ transition frequency of $\pm1.5$ GHz \cite{Balfour1972}, this corresponds to $\Gamma_L \sim W = 3$ GHz, which results in a low value of $P_\text{on}^{\text{T}}$, hence a high value of $T_\text{freq}^\text{dist}$.
An alternative is to measure different frequency points with a smaller laser linewidth and scan the whole frequency range $W$. In this scenario, the optimal value of $\Gamma_L$ must minimize the total experimental time $T_\text{exp} = T_\text{freq}N_\text{freq}$, where $N_\text{freq}$ is the number of frequency points needed to scan the range $W$. Here, one can use $N_\text{freq} = 2W/(\Gamma_L/(2\pi)) + 1$ where the laser frequency is changed in steps of $\frac{1}{2}\Gamma_L/(2\pi)$, to ensure a good overlap between the spectral shape of the laser for adjacent frequency points.
For I = 20 W/cm$^2$ the total experimental time, $T_\text{exp}$, was found to be minimized for the parameters used for Fig. \ref{signal_compare}, for which the signal depth can be resolved with 2 standard deviation confidence ($C=2$) in $T_\text{freq}=15$ minutes. With $\Gamma_L/(2\pi)= 1$ GHz we need seven frequency points and thus 1 h 45 min would be enough to cover the entire frequency range and thus obtain a signal in a reasonable amount of time. Fig. \ref{distinction} shows the simulated thermal state signal (blue) and background (black) for the same parameters as in Fig. \ref{signal_compare} including the expected one ($C=1$) and two ($C=2$) standard deviation measurement uncertainties when Eq. (\ref{criteria}) holds with equality.

\section{Discussion}\label{Discussion}

In Sec. \ref{Simu}, we discussed in detail a model to describe the expected spectroscopic signals when applying PRS in the unresolved sideband limit, where the frequency width of either the addressed transitions or the exciting light source are broader than the frequencies of the involved motional modes. As should be evident from the simulation and experimental results, this particular scenario of PRS gives rise to some very particular features. One of them is a broadening of the spectroscopic signals that can be much larger than the natural linewidth of the transition or the spectral width of the applied light source. This broadening is highly dependent on the duration of the spectroscopic light pulse and exists even in cases where the interrogation intensities are far below the relevant saturation intensity (e.g. for the case of a broad transition discussed in sub-section \ref{naturalBroad}).

At a first glance this situation may seem very unfortunate and disadvantageous with respect to applying unresolved PRS for any scientific investigations. However, in situations where the aim is to localize undetermined spectroscopic lines, this technique together with rather imprecise theoretical predictions of the line positions can turn out to be a very powerful tool to search for the transitions. More precise determination of the transition frequency can be achieved eventually by shortening the spectroscopic time and/or the frequency width of the interrogation light pulses to hone in on the specific spectral line. Clearly, in the end, the width of the spectroscopic signal will be limited either by the linewidth of the transition addressed or the intrinsic width of the light source. Having initially a narrow laser source available, actively frequency broadening it during a line search phase seems advantageous compared to just stepping the frequency of the narrow laser through a certain (large) frequency interval.    

A second important feature of unresolved sideband PRS is that the spectroscopic signal does essentially not depend on which of the two transition states are occupied by the target ion when applying the spectroscopic light pulses. In particular, in relation to localizing a narrow transition with an associated slow spontaneous decay rate, not having to initialize the target ion in the lowest lying state before applying the spectroscopic light source can speed up the time that is otherwise required to obtain a spectrum. Indeed, as in the case considered in Sec. \ref{laserBroad}, the main contribution to the spectral broadening of the intrinsically narrow line stems from broadening of the motional state distribution due to light stimulated processes, i.e., absorption and stimulated emission.

\begin{figure}[h!]      
	\begin{center}
		\includegraphics[width = 0.8\textwidth]{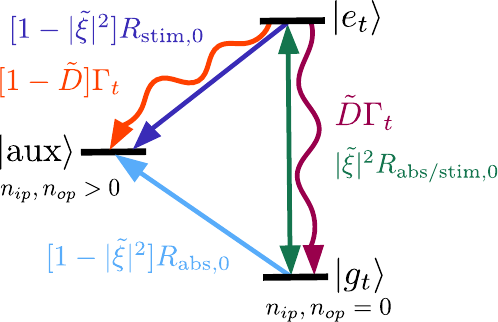}
		\caption{Level scheme of the 3-level model of PRS including the considered transition processes and their rates.}
		\label{3-level}
	\end{center}
\end{figure}

This brings us finally to a short presentation of how one can, in a simple physical picture, understand the spectroscopic signals obtained in Sec. \ref{Simu}. While it is essentially irrelevant whether the target ion is initially in the ground or excited state of the target transition, the initial motional state of the two-ion system is highly important, since the spectroscopic signal is in essence a measure of how much population is left in the motional ground state after the application of the spectroscopic light pulse. 
If the following RSB shelving pulse(s) on the readout ion could transfer all population in $n_{ip}, n_{op}>0$ to $\ket{e_r}$, the resulting fluorescence level would exactly reflect the remaining motional ground state population. From this insight, one can describe the essence of unresolved PRS through a 3-level picture as depicted in Fig. \ref{3-level}.

In this picture, in addition to the two target transition levels $\ket{g_t}$ and $\ket{e_t}$, one considers a third auxiliary level, $\ket{\text{aux}}$, which is representing the collection of all states with $n_{ip}, n_{op}>0$ and which is hence populated whenever absorption or emission leads to a change in motional state away from the ground state. During a spontaneous emission event the probability to stay within $n_{ip},n_{op} = 0$ is $\tilde{D} = D(\eta_{ip,t},\eta_{op,t},0,0,0,0)$. Since $\sum_{s_{ip},s_{op}} D(n_{ip},n_{op},s_{ip},s_{op}) = 1$ the probability to go to $\ket{\text{aux}}$ is $1 - \tilde{D}$. 
Likewise, during a single absorption or stimulated emission event the probability to stay in the motional ground state is $|\tilde{\xi}|^2 = |\xi(\eta_{ip,t},\eta_{op,t},0,0,0,0)|^2$, and since $\sum_{s_{ip},s_{op}} |\xi(n_{ip},n_{op},s_{ip},s_{op})|^2 = 1$ the probability to go to $\ket{\text{aux}}$ is $1 - |\tilde{\xi}|^2$. In principle the stimulated probabilities also depend on the specific sideband transition frequency through $R_{\text{abs/stim},0}$. However, an important feature of unresolved PRS is that the carrier and the first sideband orders are simultaneously excited by almost the same energy spectral density. This is especially the case for carrier resonance conditions, since the spectral shape of the broad transition or laser is approximately flat in the center. In the 3-level model we thus assume all transition frequencies to be equal to $\omega_t$. The rate equations for the 3-level model are thus (see Fig. \ref{3-level}):

\begin{align}
\begin{split}
&\dv{t}P_{\ket{g_t}} = -R_{\text{abs},0}P_{\ket{g_t}} + (|\tilde{\xi}|^2R_{\text{stim},0} + \tilde{D}\Gamma_t)P_{\ket{e_t}}\\
&\dv{t}P_{\ket{e_t}} = - (\Gamma_t + R_{\text{stim},0})P_{\ket{e_t}} + |\tilde{\xi}|^2R_{\text{abs},0}P_{\ket{g_t}}\\
&\dv{t}P_{\ket{\text{aux}}} =  [(1-\tilde{D})\Gamma_t + (1-|\tilde{\xi}|^2)R_{\text{stim},0}]P_{\ket{e_t}} \\
& \hspace{13.2mm}+ (1-|\tilde{\xi}|^2)R_{\text{abs},0}P_{\ket{g_t}}
\end{split}
\end{align} 
While in our full model presented in Sec. \ref{UnresolvedPRS} absorption and emission processes can also bring back population from excited motional states to the ground state, we chose to neglect it in the 3-level model for two reasons: 1) since $\ket{\text{aux}}$ is initially empty, the rates out of it will not play a role for short spectroscopy pulse times, 2) for long spectroscopic pulse times, repopulation of the motional ground state becomes negligible due to the diffusive nature of the motional state dynamics.

\begin{figure}[b]         
	\begin{center}
		\includegraphics[width = 1\textwidth]{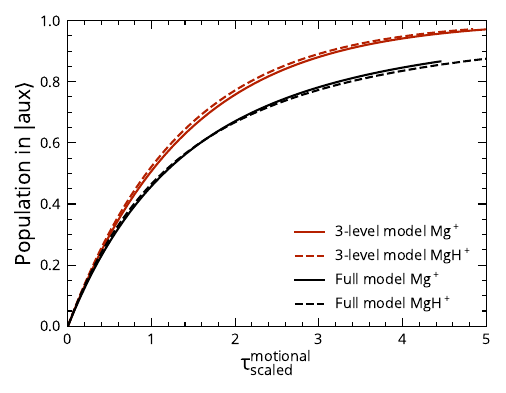}
		\vspace{-10mm}
		\caption{Comparison between the 3-level model presented in Sec. \ref{Discussion} and the full sideband model presented in Sec. \ref{UnresolvedPRS} for both the \Mg (Sec. \ref{naturalBroad}) and \MgH (Sec. \ref{laserBroad}) example cases. The population in $\ket{\text{aux}}$, representing all states with $n_{ip},n_{op}>0$, is shown as a function of the motional scaled time, which approximately corresponds to the number of excited motional quanta.}
		\label{full_3-level}
	\end{center}
\end{figure}

Since $\tilde{D} \approx |\tilde{\xi}|^2$ ($\tilde{D}=0.8608$, $|\tilde{\xi}|^2 = 0.8057$ for the \Mg case (Sec. \ref{naturalBroad}) and $\tilde{D}=0.9989$, $|\tilde{\xi}|^2 = 0.9996$ for the \MgH case (Sec. \ref{laserBroad})) PRS of systems where either spontaneous or stimulated emission dominate are not going to be fundamentally different, however the timescale of the motional ground state depletion is system dependent through $\Gamma_t$, $R_{\text{abs/stim},0}$ and $\eta_{ip}$, $\eta_{op}$. This fact can be highlighted by defining a new scaled time describing the number of motional excitations after a spectroscopy pulse time $\tau_{\text{spec}}$. Per combined absorption and emission event the probability to go to $\ket{\text{aux}}$ is $\sim 1 - |\tilde{\xi}|^4 \approx 2(\eta_{ip,t}^2+\eta_{op,t}^2)$, so we define $\tau_{\text{scaled}}^{\text{motional}} \equiv \tau_{\text{scaled}}^{t/L}\times 2(\eta_{ip,t}^2+\eta_{op,t}^2)$. Here, we naturally have to apply the relevant LDPs for the \Mg and \MgH cases giving $\eta_{ip,t}^2+\eta_{op,t}^2 = 0.22$ and $\SI{4.4e-4}{}$, respectively.
For PRS, $\tau_{\text{scaled}}^{\text{motional}}$ is the true timescale of interest, and this directly explains the large difference between the \Mg and \MgH motional populations and signal depths as a function of $\tau_{\text{scaled}}^{t/L}$. Indeed, we find that using this modified timescale for the full model results, the signal depth evolution of the two cases are (trap-induced heating aside) basically identical as seen in Fig. \ref{full_3-level} (black lines). In these examples we have $I_L = \SI{4.9}{\micro\watt\per\centi\meter\squared}$ for \Mg\, and $I_L = \SI{20}{\watt\per\centi\meter\squared}$, $\Gamma_L/(2\pi) = \SI{1}{\giga\hertz}$ for \MgH. In this figure we also show the results for the 3-level model for both the \Mg and \MgH case using the same parameters (red lines). As expected, the 3-level model gives almost identical results for the two cases, even though spontaneous emission dominates for the \Mg case, and stimulated emission dominates for the \MgH case. As expected the 3-level model follows the full model for short $\tau_{\text{scaled}}^{\text{motional}}$ and overestimates the $\ket{\text{aux}}$ population for longer times. This of course stems from the missing rates back to the motional ground state. 
Since simulating the expected spectroscopic signal by this 3-level scheme is several orders of magnitude faster than the full model presented in this paper, it can be a very useful tool as a rough guide to optimize experimental parameters.

\section{Conclusion}\label{Conclusion}

In conclusion, we have developed a model that can describe the expected spectroscopic signals when applying photon recoil spectroscopy (PRS) in the unresolved sideband limit where the linewidth of either the addressed transitions or the exciting light sources is broader than the frequencies of the involved motional modes. To test the model, we have presented experimental results with respect to the former case by carrying out unresolved sideband PRS on the 3s $^2$S$_{1/2}$ - 3p $^2$P$_{3/2}$ electronic transition of a single \Mg ion. Since very good agreement has been obtained between our experimental and simulation results, we strongly believe the model to be useful for other spectroscopic investigations. In particular, since the technique does not require initialization of the target ion in one specific state of the transition, it should be well suited to localize still vastly unknown narrow lines in various target ions, such as rovibrational transitions in molecular ions, electronic and (hyper)fine structure transitions in highly charged ions. 


\begin{acknowledgements}
S. M. and K. F. acknowledge support from the European Commission through the Marie Curie Initial Training Network COMIQ (grant agreement no 607491) under FP7. 
C. S. acknowledges support from the European Commission through the Marie Curie Individual Fellowship COMAMOC (grant agreement no 795107) under Horizon 2020. 
M. D. acknowledges support from the European Commission’s FET Open TEQ, the Villum Foundation, the Sapere Aude Initiative from the Independent Research Fund Denmark, the Danish National Laser Infrastructure, LASERLAB.DK, established by the Danish Ministry of Research and Education. This work has also been supported by the Danish National Research Foundation through the Center of Excellence ``CCQ'' (Grant agreement no.: DNRF156). 
\end{acknowledgements}


\bibliography{../../BibPRS}

\end{document}